\newif\ifconfver
\confverfalse      %declaring conference version false
\ifconfver
     \documentclass[10pt,twocolumn,twoside]{IEEEtran}
\else
    \documentclass[11pt,draftcls,onecolumn]{IEEEtran}
\fi

\usepackage{graphicx,booktabs,color}  % Written by David Carlisle and Sebastian Rahtz
\usepackage{amsmath,epsfig,amsfonts,amssymb,graphics,psfrag,theorem,calc,subfigure,url,bm,cite}
\usepackage{float}
\usepackage{stfloats}  % Written by Sigitas Tolusis
\newlength{\twidth}
\ifconfver
\else
\fi

    \makeatletter
    \def\multilimits@{\bgroup
  \Let@
  \restore@math@cr
  \default@tag
 \baselineskip\fontdimen10 \scriptfont\tw@
 \advance\baselineskip\fontdimen12 \scriptfont\tw@
 \lineskip\thr@@\fontdimen8 \scriptfont\thr@@
 \lineskiplimit\lineskip
 \vbox\bgroup\ialign\bgroup\hfil$\m@th\scriptstyle{##}$\hfil\crcr}
    \def\Sb{_\multilimits@}
    \def\endSb{\crcr\egroup\egroup\egroup}
\makeatother

\newtheorem{Lemma}{Lemma}
\newtheorem{Prop}{Proposition}
\newtheorem{Theorem}{Theorem}

\newtheorem{Corollary}{Corollary}

\theorembodyfont{\rmfamily}

{\begin{list}{}%
    {\setlength{\rightmargin}{0pc}%
    \setlength{\leftmargin}{1pc} }} %
{\end{list}}

{\begin{list}{}%
    {\setlength{\rightmargin}{1pc}%
    \setlength{\labelsep}{0pc}
    \setlength{\leftmargin}{1pc} }} %
{\end{list}}

{\begin{list}{}%
    {\setlength{\rightmargin}{0pc}%
    \setlength{\leftmargin}{0pc} }} %
{\end{list}}

{\begin{list}{}{
    \settowidth{\labelwidth}{\mbox{\textnormal{#1}}}%
    \setlength{\leftmargin}{\labelwidth+\labelsep}}}%
{\end{list}}

\begin{document}

\bibliographystyle{IEEEtran}

%
%\title{Optimal MISO Transmit Designs for Physical-Layer Secrecy using Semidefinite Programming}
\title{Optimal and Robust Transmit Designs for MISO Channel Secrecy by Semidefinite Programming}

\ifconfver \else {\linespread{1.1} \rm \fi

\author{
Qiang Li$^\dag$ and Wing-Kin Ma$^\ddag$
\thanks{$^\S$ Copyright (c) 2011 IEEE. Personal use of this material is permitted. However, permission
to use this material for any other purposes must be obtained from
the IEEE by sending a request to pubs-permissions@ieee.org. This
work is supported by a General Research Fund awarded by Research
Grant Council, Hong Kong (Project No. CUHK415908). Part of this work
appears in ICASSP 2010.}
\thanks{$^\dag$Qiang Li is with
Department of Electronic Engineering, The Chinese University of Hong
Kong, Shatin, Hong Kong S.A.R., China. E-mail: qli@ee.cuhk.edu.hk.}
\thanks{$^\ddag$Wing-Kin Ma is the corresponding author. Address:
Department of Electronic Engineering, The Chinese University of Hong
Kong, Shatin, Hong Kong S.A.R., China. E-mail: wkma@ieee.org.}
}

\maketitle

\ifconfver \else
%\begin{center} \vspace*{-2\baselineskip}
%Submitted to {\em IEEE Trans. Signal Process.} \\
%First Revision, Dec 2010 \\[2\baselineskip]
%%11th Revision, \today \\[2\baselineskip]
%\end{center}
\fi

\begin{abstract}
In recent years there has been growing interest in study of
multi-antenna transmit designs for providing secure communication
over the physical layer. This paper considers the scenario of an
intended multi-input single-output channel overheard by multiple
multi-antenna eavesdroppers. Specifically, we address the transmit
covariance optimization for secrecy-rate maximization (SRM) of that
scenario.
The challenge of this problem is that it is a nonconvex optimization problem.
%The considered SRM problem is nonconvex.
This paper shows
that the SRM problem can actually be solved in a convex and
tractable fashion, by recasting the SRM problem as a semidefinite program (SDP).
%The
%investigation mentioned above is based on the perfect channel state
%information (CSI) assumption, presently a common assumption in many
%physical-layer secrecy studies.
The SRM problem we solve is under the premise of perfect channel state information (CSI).
This paper also deals with the imperfect CSI case.
We consider a worst-case robust SRM formulation
under spherical CSI uncertainties, and we develop an optimal solution to it, again via SDP.
Moreover, our analysis reveals that
transmit beamforming is generally the
optimal transmit strategy for SRM of the considered scenario,
for both the perfect and imperfect CSI cases.
Simulation results are provided to illustrate the secrecy-rate
performance gains of the proposed SDP solutions compared to some
suboptimal transmit designs.
%The advances of multi-antenna techniques have recently led to
%renewed interest in physical-layer secrecy, a meaningful topic that
%enables us to prevent eavesdroppers from retrieving information
%intended for a legitimate user through physical layer designs. This
%paper addresses a secrecy-rate maximization problem for the scenario
%of a multi-input single-output channel listened by multiple
%multi-antenna eavesdroppers; e.g., in downlink. In the first part,
%assuming that perfect channel state information (CSI) of each link
%is known at the transmitter, we show that the respective
%secrecy-rate maximization problem, which is nonconvex, can in fact
%be turned to a convex problem, namely, a semidefinite program (SDP).
%In the second part, a deterministically-bounded model for the
%channel uncertainties is considered. We show that the respective
%robust secrecy-rate maximization problem can also be transformed
%into an SDP. Moreover, the optimal solutions of both perfect and
%imperfect CSI cases provide an important insight that transmit
%beamforming is a secrecy-rate optimal strategy for the scenario
%under consideration. Simulation results are also provided to
%illustrate that the optimal transmit design solved by our SDP
%approach can yield significantly improved secrecy rates than other
%existing closed-form designs.
\\\\
\noindent {\bfseries Index terms}$-$ Physical-layer secrecy, secrecy capacity,
transmit beamforming, semidefinite program.
\\\\
\noindent {\bfseries EDICS}: MSP-CODR (MIMO precoder/decoder design), MSP-APPL (Applications of MIMO communications and signal processing), SAM-BEAM (Applications of sensor and array multichannel processing)
\end{abstract}

%\begin{keywords}
%secrecy capacity, transmit beamforming, Charnes-Cooper
%transformation, convex optimization, semi-definite program
%(SDP).\end{keywords}
%
%%\ifconfver \else
%%   \vspace*{\baselineskip}
%%   {\bfseries EDICS}:
%%   {MSP-DECD} (MIMO space-time coding and decoding algorithms),
%%   SPC-DETC (Detection, estimation, and demodulation)
%%\fi

\ifconfver \else } \fi

\ifconfver
\else
\newpage
\fi

%==================================================

%-----------------------------------------------------------------------------

\section{Introduction}\label{introduction}

Physical-layer secrecy is an information theoretic approach where we
intend to provide a legitimate receiver with a reliable
communication, and, at the same time, make sure that illegitimate
receivers can retrieve almost nothing about the transmitted
information from the signals they have intercepted. The study of
this topic is meaningful and important, enabling us to understand
%let us gain fundamental understanding of
the information rate limits when perfect secrecy is desired; i.e.,
the secrecy capacity or the maximum secrecy rate. Moreover, the
physical-layer secrecy study provides us with vital implications on
how physical-layer secret transmit schemes should be designed in practice.
%and practical implications on how physical-layer transmit schemes should be designed for perfect secrecy.
While the concepts of physical-layer secrecy can be found back in
the 70's; e.g., the seminal works by
%{\color{blue}
Wyner~\cite{Wyner1975}, Leung-Yan-Cheong
 and
Hellman\cite{Leung1978}, and Csis\'{z}ar and
K\"{o}rner~\cite{Csiszar1978}, this topic has attracted much
interest in recent years, in both information
theory~\cite{KW2007,Khisti2007,Oggier2008,Bustin2009,Liu2009,Liang2008,Liang2007,Shafiee2007}
and signal
processing~\cite{Mukherjee09,lun2010,Negi2005,Swindlehurst2009,Zhang2009,Mukherjee2009,Jorswieck,JiaLiu2009,Li2007,JLI2009,LZHANG2009,Liao11}.
We can see at least two reasons for this. First, the rapid advances
of wireless system architectures and applications, such as those for
wireless networks, have given rise to new issues regarding
information security. In particular, the open nature of the wireless
medium means that signal interception may be easily conducted by
eavesdroppers.
%(as compared to wiretapping in wireline systems).
Cryptographic encryption, the class of techniques commonly used to
provide information security, is expected to be faced with more
challenges; for instance, in key distribution and
management~\cite{Pietro2002,CHAN2003}. Physical-layer secrecy
suggests a physical-layer-based alternative to attacking the
security problem, which is meaningful and may complement
%cryptographic encryption (which is network-layer-based).
the network-layer-based cryptographic encryption techniques.
%The study of this topic also lets us gain fundamental understanding of the information rate limit when perfect secrecy is desired.
Second, multi-input multi-output (MIMO) techniques provide
physical-layer secrecy with new and exciting opportunities.
Intuitively, if we know the channel state information (CSI) of the
eavesdroppers to a certain extent (say, in scenarios where the
eavesdroppers are users of the system, who attempt to access
unauthorized services), then we may utilize the MIMO degree of
freedom to weaken these eavesdroppers' receptions.
Another idea that is possible only with MIMO is that of interfering eavesdroppers through artificially generated spatial noise;
see \cite{Negi2005} for the original work and \cite{Mukherjee09,Jorswieck,Swindlehurst2009,Mukherjee2009,Liao11} for some recent developments.

In fact, we have recently seen a growing body of physical-layer
secrecy literatures that deal with various MIMO scenarios. For the
scenario of an MIMO channel overheard by one multi-antenna
eavesdropper, the secrecy capacity has been considered
in~\cite{KWWE2007,Khisti2007,Oggier2008,Bustin2009,Liu2009}. The
MISO counterpart has also caught some attention;
see~\cite{Shafiee2007,KW2007}. We should note that there is a
difference between the above described MIMO and MISO scenarios, from
a viewpoint of transmit optimization. Specifically, we are
interested in
%the aspects of
transmit covariance designs for achieving the maximum secrecy rate.
%Specifically, we are concerned with
%transmit covariance optimization problems for secrecy-rate
%maximization (SRM).
This secrecy-rate maximization (SRM) problem is nonconvex, and solving it
under the general MIMO scenario is a challenging issue;
see\cite{JLI2009,JiaLiu2009} for some recent endeavors.
One exception where the SRM problem is found to be tractable is
%when
%the eavesdropper has only one antenna;
%%. In that special case,
%the SRM problem in that case
%can be solved by the quasi-convex optimization
%approach in~\cite{Zhang2009} \footnote{We should note that the
%quasi-convex optimization approach~\cite{Zhang2009} can be applied
%to the case of multiple, single-antenna, eavesdroppers.}. Another
%exception is the MISO scenario
the MISO scenario
--- its optimal transmit design is
shown to admit a closed-form solution, involving a generalized
eigenvalue problem~\cite{KW2007}.
%That optimal transmit covariance
%has also been found to have a rank-one structure (excluding the
%trivial case of zero secrecy rate), implying that transmit
%beamforming is generally the optimal transmit strategy for
%%MISO, one-eavesdropper, SRM.
%the SRM of a MISO channel overhead by one multi-antenna
%eavesdropper.

This paper considers the scenario of an MISO channel overheard by
{\it multiple} multi-antenna eavesdroppers. This scenario would happen,
say, for example, in a downlink environment where the intended
receiver uses one antenna for low operational costs and the
eavesdroppers are users of the same network employing sophisticated
multi-antenna hardware to improve their interceptions. The SRM
problem for this particular scenario has been formulated
in~\cite{Liang2007}. The problem is again nonconvex, and, to our knowledge,
%the solution to the corresponding problem has
an efficient optimal transmit solution to it has not been available in general.
%{\color{blue}For the case of one-antenna Eves, \cite{Zhang2009}
%gives the optimal design by relating the SRM with the cognitive
%radio design, while for the case of multiple multi-antenna Eves,
%only the upper bound and the lower bound on the secrecy rate are
%available~\cite{Zhang2009}.}
We should mention that for the special case of one-antenna eavesdroppers,
an optimal SRM design has been proposed in \cite{Zhang2009};
the idea there is to establish a relationship between the SRM problem and
a cognitive radio (CR) design problem under the one-antenna eavesdroppers/secondary-users context.
That secrecy-CR relationship result is applicable also to MIMO intended channel.
For the multiple-antenna eavesdroppers case,
the secrecy-CR relationship result does not apply
and it is used as an approximation to the SRM problem \cite{Zhang2009}.
%{\color{blue} except for some suboptimal designs dealing with the
%upper bound and lower bound of this SRM problem~\cite{Zhang2009}.}
This paper %addresses this problem, showing
shows
that the considered SRM
problem can actually be solved in a convex and tractable fashion;
specifically, we turn the SRM problem to a
semidefinite program (SDP), a representative convex optimization
problem whose globally optimal solution can be obtained efficiently
by available algorithms~\cite{Grant2009,Sturm1999}.
%To obtain this desirable result is indirect,
The proposed approach is indirect, and it may remind us of study of
some kind of dual problems\footnote{The dual problem relationship we refer to is that of the quality-of-service (QoS) constrained power minimization problem and the power constrained QoS maximization problem \cite{Wiesel06}.}
in the context of
multiuser downlink transmit beamforming \cite{Wiesel06,Schubert04};
%{\color{red} {\it (take out Schubert's reference since the duality relationship was not shown there?)}}
also \cite{Jorswieck09,Wunder08} for multiuser downlink OFDM.
%the context of multiuser transmit optimization \cite{Jorswieck09,Wiesel06,Schubert04,Seong06,Wunder08}:
In particular, we need to consider another secrecy-rate formulation, namely, a secrecy-rate constrained
problem, and use the results established there to prove that the SRM
problem has an SDP equivalent.

%In particular, we need to study
%another secrecy-rate formulation, namely, a secrecy-rate constrained
%problem, and use the results established there to prove that the SRM
%problem has an SDP equivalent. {\color{blue}Exploiting this kind of
%duality in downlink transmit design can also be found in
%\cite{Jorswieck09,Wiesel06,Schubert04,Seong06,Wunder08}.}
%%As a side product of
%%solving SRM, we build an SDP solution to the secrecy-rate
%%constrained problem, which is also an interesting secrecy transmit
%%design.

Another key contribution of this work is that we extend our SRM
results to the imperfect CSI case. The eavesdroppers' CSIs may be
estimated depending on the application environment; e.g., when the
eavesdroppers are also system users,
%but it is not always possible to keep an accurate track of those CSIs.
however we may only have some inaccurate, possibly rough, knowledge
about those CSIs. The CSI of the intended receiver may also be
subject to some uncertainties; e.g., those caused by channel
estimation errors and/or quantization errors, though such an issue
should be less severe than that for the eavesdroppers. We should
note that physical-layer secrecy with imperfect CSI or no CSI has
started to catch attention very recently; see, e.g.,
%\cite{Jorswieck,Swindlehurst2009,KW2007,LZHANG2009,Mukherjee10,Wolf2010}.
\cite{Jorswieck,Swindlehurst2009,KW2007,LZHANG2009,Mukherjee10}.
Here, we consider a robust SRM formulation of our considered
scenario,
%for the scenario of a MISO channel overheard by multiple multi-antenna eavesdroppers,
where
we employ a spherical CSI uncertainty model
and the uncertainties are handled in a worst-case sense.
%the CSI uncertainties
%are handled in a worst-case sense.
The proposed robust SRM formulation is a conservative design---
it guarantees perfect secrecy for any admissible CSI uncertainties, including the worst.
%It is noteworthy that conservative designs are meaningful when dealing with slowly fading channels.
The robust SRM formulation
exhibits a more complex structure than its non-robust counterpart.
Fortunately, it is shown that the robust SRM problem can be
equivalently represented, and solved, by an SDP; in addition to the
approach used in non-robust SRM, the idea is to employ linear matrix
inequality characterization to handle the imperfect-CSI induced
constraints in a tractable way.

%In establishing our optimal SDP solutions to the SRM problems,
%we find an interesting physical result that agrees with intuitive expectation---
%for the considered scenario,
%%(MISO channel overheard by multiple multi-antenna eavesdroppers),
%transmit beamforming is generally the SRM optimal transmit strategy for both the perfect and imperfect CSI cases.
%While this result is already known for the one multi-antenna eavesdropper case~\cite{KW2007} and the multiple one-antenna eavesdroppers case~\cite{Zhang2009}, with perfect CSI,
%the result here is more general.

In establishing our optimal SDP solutions to the SRM problems,
we obtain a physical result that agrees  well with intuitive expectation ---
transmit beamforming is generally the SRM optimal transmit strategy of the considered scenario, for both the perfect and imperfect CSI cases.
While this result is already known for the one multi-antenna eavesdropper case~\cite{KW2007} and the multiple one-antenna eavesdroppers case~\cite{Zhang2009} (both with perfect CSI),
the result here is more general.

This paper is organized as follows. A background review and problem
statement is given in Section~\ref{background}. Section~\ref{SRM}
considers the SRM problem for the MISO, multi-eavesdropper
scenario with perfect CSI, and establishes an SDP solution to it.
Section~\ref{Robust_SRM} extends the SRM results to the imperfect
CSI case. Simulation results comparing the proposed SRM solutions
and some other suboptimal secrecy transmit designs are illustrated
in Section\ref{simulation}. Section~\ref{conclusion} gives the
conclusion and discussion.

%--------Notations---------------
Our notations are as follows. We will use boldface capital letters
to denote matrices, boldface lower case letters to denote vectors;
$\mathbf{A}^T$, $\mathbf{A}^H$,
${\bf A}^{\dag}$,
$\text{Tr}(\mathbf{A})$ and $\det(\mathbf{A})$ represent transpose,
Hermitian (conjugate) transpose,
pseudo inverse,
trace and
determinant of a matrix $\mathbf{A}$;
%$\mathbf{I}_N$ is the $N$-by-$N$ identity matrix;
$\mathbf{I}$ denotes an identity matrix; $\|\cdot\|_F$ represents
the Frobenius norm of a matrix; rank(${\bf A}$) is the rank of
matrix ${\bf A}$; $\mathbf{A}\succeq \mathbf{0}$ $({\bf A}\succ {\bf
0})$ means $\mathbf{A}$ is a Hermitian positive semidefinite
(definite) matrix; $\mathbb{R}_{+}$ denotes the set of all nonnegative
real numbers;
$\mathbb{H}^{N}$ denotes the set of all $N$-by-$N$ Hermitian matrices;
$\mathbb{H}_{+}^{N}$ denotes the set of all $N$-by-$N$ Hermitian positive semidefinite matrices;
$\mathbb{C}^{M \times N}$ represents an $M$-by-$N$ dimensional
complex matrix set; ${\bf A}\otimes {\bf B}$ denotes the Kronecker
product of ${\bf A}$ and ${\bf B}$; ${\rm vec}({\bf A})$ denotes the
vectorization of matrix ${\bf A}$ by stacking its columns;
$\text{E}\{\cdot\}$ is the expectation operator.

\section{Background and Problem Statement}\label{background}

In this section we review some basic concepts of physical-layer
secrecy with an emphasis on the MISO scenarios, and provide the
problem statement of our interested scenario.

\subsection{One Eavesdropper Case: A Review}\label{background_OneEve}

\begin{figure}[htp]
\begin{center}
\subfigure[][]{\resizebox{.28\textwidth}{!}{\includegraphics{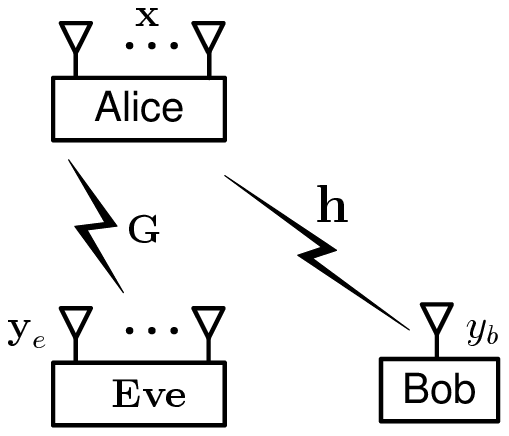}}}
\subfigure[][]{\resizebox{.35\textwidth}{!}{\includegraphics{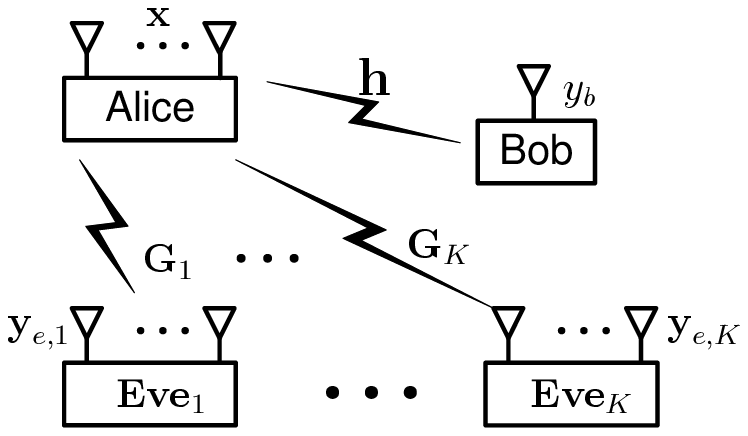}}}
\end{center}
\caption{
System model.
(a) One multi-antenna eavesdropper;
(b) Multiple multi-antenna eavesdroppers.
}
\label{fig:example0}
\end{figure}

The endeavor of this subsection is to give some intuitive insights
into physical-layer secrecy, by reviewing the scenario of a MISO
channel overheard by one eavesdropper (also known as the MISOME
wiretap channel in the literature~\cite{KW2007}).
Fig.~\ref{fig:example0}(a) shows the scenario. In this
configuration, the transmitter is equipped with multiple antennas,
intending to deliver information to the legitimate receiver which
uses one antenna. One eavesdropper is present, and it uses multiple
antennas to intercept the transmission. Following the convention in
the secrecy capacity literature, we call the transmitter, the
legitimate receiver, and the eavesdropper as {\it Alice}, {\it Bob},
and {\it Eve}, respectively. The signal models for the Alice-to-Bob
and Alice-to-Eve links are, respectively,
\begin{subequations}\label{eq:MISOME_model}
\begin{align}
y_b(t) & = {\bf h}^H {\bf x}(t) + n(t), \label{eq:MISOME_model_a}\\
{\bf y}_e(t) & =  {\bf G}^H {\bf x}(t) + {\bf
v}(t).\label{eq:MISOME_model_b}
\end{align}
\end{subequations}
Here, ${\bf x}(t) \in \mathbb{C}^{N_t}$ is the transmit vector by
Alice, with $N_t$ being the number of transmit antennas; ${\bf h}
\in \mathbb{C}^{N_t}$ represents the MISO Alice-to-Bob channel;
${\bf G} \in \mathbb{C}^{N_t \times N_{e}}$ represents the MIMO
Alice-to-Eve channel, with $N_e$ being the number of receive
antennas at Eve; and $n(t) \in \mathbb{C}$, ${\bf v}(t) \in
\mathbb{C}^{N_e}$ are zero-mean additive white Gaussian
noises. Without loss of generality, we assume unit variance of all
the noise terms; i.e., ${\rm E}\{ |n(t)|^2 \} = 1$, ${\rm E}\{ {\bf
v}(t) {\bf v}^H(t) \} = {\bf I}$.

The problem of interest here is to prevent Eve from retrieving
useful information via appropriate transmit designs, and this can be
addressed by using the notion of physical-layer
secrecy~\cite{LiangPoor2008}. To describe this, let us denote the
transmit covariance by
\[ {\bf W} = {\rm E}\{ {\bf x}(t) {\bf x}^H(t) \}. \]
According to~\cite{KW2007}, the transmit covariance design that
provides the maximum secrecy rate for the system
model~\eqref{eq:MISOME_model} is given by
\begin{equation}\label{eq:SRM_MISOME}
\begin{array}{rl}
R^\star(P) = \displaystyle \max_{\bf W} &  \log( 1 + {\bf h}^H {\bf W} {\bf h}) - \log\det( {\bf I} + {\bf G}^H {\bf W} {\bf G})  \\
{\rm s.t.} &  {\bf W} \succeq {\bf 0}, ~ {\rm Tr}({\bf W}) \leq P,
\end{array}
\end{equation}
where $P$ is a given average transmit power limit, and $R^\star(P)$
is defined to be the optimal secrecy rate (in bps/Hz) for a given
$P$. As seen in~\eqref{eq:SRM_MISOME}, the problem is to maximize
the mutual information difference between the Alice-to-Bob and
Alice-to-Eve channels. The subsequent optimized rate $R^\star(P)$ is
achievable--- from an information theoretic perspective, there exist
codes such that Bob can obtain a perfectly secure message from Alice
at a rate of $R^\star(P)$ bps/Hz, while Eve can retrieve almost
nothing about the message~\cite{Csiszar1978}. It has also been shown
that~\eqref{eq:SRM_MISOME} is the secrecy capacity for the one-Eve
model in~\eqref{eq:MISOME_model}.

The secrecy-rate maximization problem in~\eqref{eq:SRM_MISOME} has a
closed-form solution. Let ${\bf q}$ be the unit-norm principal
generalized eigenvector
%~\cite{Golub1996}
of
$( {\bf I} + P{\bf h}{\bf h}^H, {\bf I} + P{\bf G}{\bf G}^H )$.
The optimal transmit design of~\eqref{eq:SRM_MISOME}, denoted herein
by ${\bf W}^\star$, can be shown to be~\cite{KW2007}
\begin{equation}\label{solution_MISOME}
{\bf W}^\star = \left\{
\begin{array}{rl}
P {\bf q} {\bf q}^H, & f(P {\bf q} {\bf q}^H) > 0 \\
{\bf 0}, & f(P {\bf q} {\bf q}^H) \leq 0
\end{array} \right.
\end{equation}
where $f({\bf W})= \log( 1 + {\bf h}^H {\bf W} {\bf h}) - \log\det(
{\bf I} + {\bf G}^H {\bf W} {\bf G})$ denotes the secrecy rate of a
given ${\bf W}$.
The structure of ${\bf W}^\star$ shown above
reveals two physical results: First, the optimal transmit
strategy for the case of $R^\star(P) > 0$ (positive secrecy rate) is
to employ a rank-one transmit structure; i.e., transmit beamforming.
Second, the optimal transmit design for the (trivial) case of
$R^\star(P) = 0$ is to shut down the transmitter.

%ken_17-06-2010:
%
%The secrecy-rate maximization problem in~\eqref{eq:SRM_MISOME} has a
%closed-form solution. Let ${\bf u}$ be the unit-norm principal
%generalized eigenvector~\cite{Golub1996} of
%\[ ( {\bf I} + P{\bf h}{\bf h}^H, {\bf I} + P{\bf G}{\bf G}^H ). \]
%The optimal transmit design of~\eqref{eq:SRM_MISOME}, denoted herein
%by ${\bf W}^\star$, can be shown to be~\cite{KW2007}
%\begin{equation}\label{solution_MISOME}
%{\bf W}^\star = \left\{
%\begin{array}{rl}
%P {\bf u} {\bf u}^H, & f(P {\bf u} {\bf u}^H) > 0 \\
%{\bf 0}, & f(P {\bf u} {\bf u}^H) \leq 0
%\end{array} \right.
%\end{equation}
%where $f({\bf W})= \log( 1 + {\bf h}^H {\bf W} {\bf h}) - \log\det(
%{\bf I} + {\bf G}^H {\bf W} {\bf G})$ denotes the secrecy rate of a
%given ${\bf W}$.
%The structure of ${\bf W}^\star$ shown above
%reveals two interesting results: First, the optimal transmit
%strategy for the case of $R^\star(P) > 0$ (positive secrecy rate) is
%to employ a rank-one transmit structure; i.e., transmit beamforming.
%Second, the optimal transmit design for the (trivial) case of
%$R^\star(P) = 0$ is to shut down the transmitter.

\subsection{Multi-Eavesdropper Case}\label{background_MultiEve}

This paper focuses on the scenario of MISO Alice-to-Bob link and
multiple MIMO Alice-to-Eves links; such a configuration is depicted in Fig.~\ref{fig:example0}(b).
For this scenario the system model is modified to
\begin{subequations}\label{eq:MISOMEs_model}
\begin{align}
y_b(t) & = {\bf h}^H {\bf x}(t) + n(t), \label{eq:MISOMEs_model_a} \\
{\bf y}_{e,k}(t) & =  {\bf G}^H_k {\bf x}(t) + {\bf v}_k(t), \quad
k=1,\ldots,K,\label{eq:MISOMEs_model_b}
\end{align}
\end{subequations}
where the Alice-to-Bob channel model in~\eqref{eq:MISOMEs_model_a}
is identical to that in~\eqref{eq:MISOME_model_a}; ${\bf
y}_{e,k}(t)$ is the received signal at the $k$th Eve; ${\bf G}_k \in
\mathbb{C}^{N_t \times N_{e,k}}$ is the MIMO channel from Alice to
the $k$th Eve; $N_{e,k}$ is the number of receive antennas of the
$k$th Eve;
${\bf v}_k(t)$ is additive white Gaussian noise with zero mean and ${\rm E}\{ {\bf v}_k(t) {\bf v}_k(t)^H \} = {\bf I}$;
$K$ is the number of Eves. An achievable secrecy rate
for the multiple-Eve model~\eqref{eq:MISOMEs_model} has been derived
in~\cite{Liang2007} and is given by
\begin{equation}\label{SRM_MISOMEs}
\begin{array}{rl}
R^\star(P) = \displaystyle \max_{\bf W} & \displaystyle \min_{k=1,\ldots,K} f_k({\bf W}) \\
{\rm s.t.} &  {\bf W} \succeq {\bf 0}, ~ {\rm Tr}({\bf W}) \leq P
\end{array}
\end{equation}
where
\begin{equation}\label{mutual_inf_diff}
f_k({\bf W})=  \log( 1 + {\bf h}^H {\bf W} {\bf h}) - \log\det( {\bf
I} + {\bf G}_k^H {\bf W} {\bf G}_k)
\end{equation}
is the secrecy rate function corresponding to the $k$th Eve. The
goal of~\eqref{SRM_MISOMEs} is to maximize the worst secrecy rate
(or mutual information difference) among all the Eves.
%We should mention that $R^\star(P)$ in (5), the maximum worst-case secrecy rate, is achievable.

The secrecy-rate maximization (SRM) problem~\eqref{SRM_MISOMEs}
presents a challenge from the standpoint of transmit covariance
optimization. Problem~\eqref{SRM_MISOMEs} is a nonconvex problem,
due to the nonconcave, Eve-induced terms $-\log\det( {\bf I} + {\bf
G}_k^H {\bf W} {\bf G}_k)$. While this nonconvex SRM problem admits
a closed-form solution in the one-Eve case (as mentioned above),
%this is not so with the multi-Eve case {\color{blue} in general}.
it is not known if the SRM problem can also be solved in closed form
for the general multi-Eve case.

We should mention a transmit design that has a closed-form solution
but is generally suboptimal in the SRM context, namely, {\it
projected maximum-ratio transmission (projected-MRT)}. The idea is
to apply nulling on all Eves; i.e., to enforce ${\bf G}_k^H {\bf W}=
{\bf 0}$ for all $k$. Let $\bm{G}= [~ {\bf G}_1, \ldots, {\bf G}_K
~]$ be the aggregate channel matrix of all Eves, and
$\bm{\Pi}_{\bm{G}}^\perp = {\bf I} - \bm{G} ( \bm{G}^H \bm{G})^\dag
\bm{G}^H$ be the orthogonal complement projector of $\bm{G}$. The
transmit beamformer weight in projected-MRT is
\[
{\bm w} = \frac{\sqrt{P} }{ \| \bm{\Pi}_{\bm{G}}^\perp {\bf h} \|_2
} \bm{\Pi}_{\bm{G}}^\perp {\bf h},
\]
or, in other words, we choose ${\bf W}= {\bm w} {\bm w}^H$. The
advantage of projected-MRT lies in its simplicity, but its downside
is that we may not have the degree of freedom to perform nulling
when the total number of antennas of Eves, $\sum_{k=1}^K N_{e,k}$,
reaches or exceeds the number of transmit antennas $N_t$.

In what follows, we will describe our approach to solving the SRM
problem~\eqref{SRM_MISOMEs}.
% in a globally optimal fashion.

%---------------------------------------------------------------------------
\section{An SDP Approach to the Secrecy Rate Maximization
Problem}\label{SRM}

The focus of this section is on solving Problem~\eqref{SRM_MISOMEs},
the secrecy-rate maximization of an MISO channel eavesdropped by
multiple multi-antenna eavesdroppers. We will show how the SRM
problem~\eqref{SRM_MISOMEs}, which is nonconvex, can actually be
solved by an equivalent SDP problem that is convex and tractable.
%In the process of our development, we are also able to assert that transmit beamforming is generally the optimal transmit strategy for (5).

The idea behind the proposed SDP solution is to consider some form
of convex approximation to the SRM~\eqref{SRM_MISOMEs}, and then to
prove that that approximate SRM problem is indeed tight. The latter
part is not straightforward. We will need to study a variation of
SRM, given as follows:
\begin{equation}\label{PM_SRC}
\begin{array}{rl}
P^\star(R) = \displaystyle \min_{{\bf W} \succeq {\bf 0}} &  {\rm Tr}({\bf W}) \\
{\rm s.t.} & \displaystyle \min_{k=1,\ldots,K} f_k({\bf W}) \geq R.
\end{array}
\end{equation}
The design in~\eqref{PM_SRC} seeks to satisfy a minimum secrecy rate
specification $R$ (which is given) and to minimize the average
power. The reason for considering the secrecy-rate constrained (SRC)
problem~\eqref{PM_SRC} is that it is relatively easier to analyze
than the SRM~\eqref{SRM_MISOMEs}. And, as a side advantage, the SRC
design itself is interesting and practically meaningful. We will
develop some key results for Problem~\eqref{PM_SRC}, e.g.,
that~\eqref{PM_SRC} has an SDP equivalent. We will then establish a
link between the SRC problem~\eqref{PM_SRC} and the SRM
problem~\eqref{SRM_MISOMEs}, thereby allowing us to use the proven
results in SRC to show that SRM can be exactly solved by an SDP.

The first and second subsections describe our developments for the
SRC and SRM problems, respectively.

\subsection{The Secrecy-Rate Constrained Problem}\label{SRM_SRC}

To study the SRC problem~\eqref{PM_SRC}, let us
express~\eqref{PM_SRC} in a more explicit form:
\begin{equation}\label{PM_SRC_reform}
\begin{array}{rl}
P^\star(R) = \displaystyle \min_{{\bf W} \succeq {\bf 0}} &  {\rm Tr}({\bf W}) \\
{\rm s.t.} &
%\displaystyle \min_{k=1,\ldots,K} \frac{1 + {\bf h}^H {\bf W} {\bf h}}{ \det( {\bf I} + {\bf G}_k^H {\bf W} {\bf G}_k ) } \geq 2^R
\displaystyle 2^{-R} \geq \max_{k=1,\ldots,K} \frac{ \det( {\bf I} +
{\bf G}_k^H {\bf W} {\bf G}_k ) }{ 1 + {\bf h}^H {\bf W} {\bf h}  }.
\end{array}
\end{equation}
%where we have substituted the functional form of $f_k(\cdot)$ [in (6)] into Problem (7).
Problem~\eqref{PM_SRC_reform} is nonconvex, due to the determinant
functions in the constraint of~\eqref{PM_SRC_reform}. Consider the
following lemma which we will use to provide a convex approximation
to~\eqref{PM_SRC_reform}:

\begin{Lemma} Let ${\bf A} \succeq {\bf 0}$.
%, ${\bf A} \neq {\bf 0}$.
It holds true that
\begin{equation}\label{det_trace_ineq}
\det( {\bf I}+ {\bf A} ) \geq 1 + {\rm Tr}({\bf A}),
\end{equation}
and that the equality in~\eqref{det_trace_ineq} holds if and only if
%${\bf A}$ is of rank one.
${\rm rank}({\bf A}) \leq 1$.
\end{Lemma}

{\it Proof:} \
Let $r= {\rm rank}({\bf A})$.
The case of $r= 0$ is trivial.
For $r \geq 1$,
let $\lambda_1 \geq \lambda_2 \geq \ldots \geq
\lambda_r
> 0$ denote the non-zero eigenvalues of ${\bf A}$.
We have that
\begin{align*}
\det({\bf I} + {\bf A}) = \textstyle \prod_{i=1}^r (1+ \lambda_i) & \textstyle  = 1 + \sum_{i=1}^r \lambda_i + \sum_{i \neq k} \lambda_i \lambda_k + \ldots \\
& \geq \textstyle 1 + \sum_{i=1}^r \lambda_i = 1 + {\rm Tr}({\bf A})
\end{align*}
and it can be seen that the equality above holds if and only if $r=
1$.\hfill $\blacksquare$

Applying Lemma~1 to Problem~\eqref{PM_SRC_reform}, we obtain a
relaxation of~\eqref{PM_SRC_reform} as follows:
\begin{equation}\label{RPM_SRC}
\begin{array}{rl}
P^\star(R) \geq P^\star_{relax}(R) = \displaystyle \min_{{\bf W} \succeq {\bf 0}} &  {\rm Tr}({\bf W}) \\
{\rm s.t.} & \displaystyle 2^{-R} \geq \max_{k=1,\ldots,K} \frac{ 1
+ {\rm Tr}({\bf G}_k^H {\bf W} {\bf G}_k ) }{ 1 + {\bf h}^H {\bf W}
{\bf h}  }.
%{\rm s.t.} & \displaystyle \min_{k=1,\ldots,K} \frac{1 + {\bf h}^H {\bf W} {\bf h}}{ 1 + {\rm Tr}({\bf G}_k^H {\bf W} {\bf G}_k ) } \geq 2^R
\end{array}
\end{equation}
Problem~\eqref{RPM_SRC} is a convex problem. Specifically,
\eqref{RPM_SRC} can be formulated as an SDP
\begin{equation}\label{RPM_SRC_SDP}
\begin{array}{rl}
P^\star_{relax}(R) = \displaystyle \min_{{\bf W} } &  {\rm Tr}({\bf W}) \\
{\rm s.t.} & {\bf W} \succeq {\bf 0}, \\
& 1 + {\rm Tr}({\bf h} {\bf h}^H {\bf W})  \geq 2^R( 1 + {\rm Tr}( {\bf G}_k{\bf G}_k^H {\bf W}  ) ), \\
& \quad k=1,\ldots,K,
\end{array}
\end{equation}
whose globally optimal solution can be efficiently found by
available solvers~\cite{Grant2009,Sturm1999}. While our original
motivation is to approximate the SRC problem~\eqref{PM_SRC_reform}
by the convex relaxation~\eqref{RPM_SRC}, Lemma~1 provides an
important hint regarding the tightness of the relaxation: The
relaxation~\eqref{RPM_SRC} is tight (which means $P^\star(R) =
P^\star_{relax}(R)$) when the optimal solution of~\eqref{RPM_SRC} is
of rank one. We prove the following key result:

\begin{Prop}\label{prop:PM_rank1}
Consider the relaxed SRC problem~\eqref{RPM_SRC} for the case where
the secrecy-rate specification $R$ is positive\footnote{The case of
$R= 0$ is trivial since ${\bf W}= {\bf 0}$ can easily be verified to
be the corresponding optimal solution of~\eqref{RPM_SRC}.}, or
simply $R > 0$. Also, suppose that Problem~\eqref{RPM_SRC} is
feasible. Then, the optimal solution of~\eqref{RPM_SRC} must be of
rank one and unique.
%Suppose that the relaxed SRC problem~\eqref{RPM_SRC} is
%{\color{blue}feasible for $R
%> 0$}. Then, the optimal solution of~\eqref{RPM_SRC}
%must be of rank one and unique\footnote{The case of $R= 0$ is
%trivial since ${\bf W}= {\bf 0}$ can easily be verified to be the
%corresponding optimal solution of~\eqref{RPM_SRC}.}.
\end{Prop}

The proof, which can be found in Appendix A, is based on
examination of the Karush-Kuhn-Tucker (KKT) conditions of the
SDP~\eqref{RPM_SRC_SDP}.

Using Proposition 1 and Lemma 1, we reach the conclusion that
\begin{Corollary}
Consider the SRC problem~\eqref{PM_SRC_reform} for the case $R > 0$,
and suppose that Problem~\eqref{PM_SRC_reform} is feasible.
%Suppose that the SRC problem~\eqref{PM_SRC_reform} is
%{\color{blue}feasible for $R
%> 0$}.
The relaxed SRC problem~\eqref{RPM_SRC} [or the SDP
~\eqref{RPM_SRC_SDP}] exactly solves the SRC problem, in the sense
that the optimal solution of~\eqref{RPM_SRC} is also that
of~\eqref{PM_SRC_reform}, and vice versa. Moreover, the optimal SRC
solution is unique and of rank one.
\end{Corollary}

%\begin{proof}
{\it Proof:} \ Let $\hat{\bf W}$ be the optimal solution
of~\eqref{RPM_SRC}, which is unique and of rank one by Proposition
1. By Lemma~1, $\hat{\bf W}$ is feasible to~\eqref{PM_SRC_reform},
which implies that $P^\star_{relax}(R) = {\rm Tr}(\hat{\bf W}) \geq
P^\star(R)$. Since we also have $P^\star(R) \geq P^\star_{relax}(R)$
[cf., Eqn.~\eqref{RPM_SRC}], we conclude that $P^\star_{relax}(R) =
P^\star(R)$; i.e., $\hat{\bf W}$ is optimal
to~\eqref{PM_SRC_reform}.

On the other hand, let ${\bf W}^\star$ be an optimal solution of
~\eqref{PM_SRC_reform}. Owing to Lemma 1, ${\bf W}^\star$ is
feasible to~\eqref{RPM_SRC}. The result $P^\star_{relax}(R) =
P^\star(R) = {\rm Tr}({\bf W}^\star)$ further implies that ${\bf
W}^\star$ is optimal to~\eqref{RPM_SRC}, too. By Proposition 1,
${\bf W}^\star$ has to be unique and of rank one.\hfill
$\blacksquare$\\

We have developed an SDP solution to the SRC
problem~\eqref{PM_SRC_reform}.
Moreover, the unique rank-one nature of the SRC solution, as indicated in Corollary~1,
suggests that
transmit beamforming
is generally the optimal transmit strategy for the SRC
problem~\eqref{PM_SRC_reform}.

\subsection{The Secrecy-Rate Maximization Problem}\label{SRM_SRM}

We now turn our attention to the SRM problem~\eqref{SRM_MISOMEs}.
For convenience, we rewrite Problem~\eqref{SRM_MISOMEs} as
\begin{equation}\label{SRM_MISOMEs_reform}
\gamma^\star(P) = \min_{ \substack{ {\bf W} \succeq {\bf 0} \\ {\rm
Tr}({\bf W}) \leq P } }   \max_{k=1,\ldots,K} \frac{ \det( {\bf I} +
{\bf G}_k^H {\bf W} {\bf G}_k )}{ 1 + {\bf h}^H {\bf W} {\bf h}}
\end{equation}
where $0 < \gamma^\star(P) \leq 1$ is related to the optimal secrecy
rate through the relation $R^\star(P) = \log(1/\gamma^\star(P))$.
Following the same spirit as in the preceding subsection, we apply
Lemma 1 to~\eqref{SRM_MISOMEs_reform} to obtain a relaxation:
\begin{equation}\label{Relx_SRM_MISOMEs}
\gamma^\star(P) \geq  \gamma^\star_{relax}(P) = \min_{ \substack{
{\bf W} \succeq {\bf 0} \\ {\rm Tr}({\bf W}) \leq P } }
\max_{k=1,\ldots,K} \frac{ 1 + {\rm Tr}({\bf G}_k^H {\bf W} {\bf
G}_k )}{ 1 + {\bf h}^H {\bf W} {\bf h}}.
\end{equation}
The relaxation above is a quasi-convex problem, whose globally
optimal solution can be searched by general techniques for
quasi-convex optimization (e.g., bisection \cite{Boyd2004}). We
however will propose a more efficient method of
solving~\eqref{Relx_SRM_MISOMEs}; namely, via SDP, after we study
the tightness of the relaxation in~\eqref{Relx_SRM_MISOMEs}.

Our key question here is whether Problem~\eqref{Relx_SRM_MISOMEs}
yields a rank-one solution: If it does, then the
relaxation~\eqref{Relx_SRM_MISOMEs} is tight ($\gamma^\star(P) =
\gamma^\star_{relax}(P)$) by Lemma 1. Our insight to this is as
follows: Suppose that $\gamma^\star_{relax}(P)$
in~\eqref{Relx_SRM_MISOMEs} has been computed. We consider the
following relaxed SRC problem
\begin{equation}\label{Relx_PM_SRC}
\begin{array}{rl}
\displaystyle \min_{{\bf W} \succeq {\bf 0} } &  {\rm Tr}({\bf W}) \\
{\rm s.t.} & \gamma^\star_{relax}(P) \geq \displaystyle
\max_{k=1,\ldots,K} \frac{ 1 + {\rm Tr}({\bf G}_k^H {\bf W} {\bf
G}_k )}{ 1 + {\bf h}^H {\bf W} {\bf h}},
\end{array}
\end{equation}
where we fix the secrecy rate specification at $R=
\log(1/\gamma^\star_{relax}(P))$. By Proposition 1, we know that
Problem~\eqref{Relx_PM_SRC} has a unique rank-one solution (in the
nontrivial case). If the optimal solution of~\eqref{Relx_PM_SRC} can
be proven to be an optimal solution of~\eqref{Relx_SRM_MISOMEs} as
well, then we will be able to infer immediately that the
relaxation~\eqref{Relx_SRM_MISOMEs} is tight. With this idea in
mind, we give a formal proof in Appendix B and show that

\begin{Theorem}\label{Theorem:SRM_rank1}
Consider the SRM problem~\eqref{SRM_MISOMEs_reform} for the case of
$0< \gamma^\star(P) < 1$. The relaxed SRM
problem~\eqref{Relx_SRM_MISOMEs} exactly solves the SRM problem, in
the sense that the optimal solution of~\eqref{Relx_SRM_MISOMEs} is
also that of~\eqref{SRM_MISOMEs_reform}, and vice versa. Moreover,
the optimal SRM solution is unique and of rank one\footnote{The case
of $\gamma^\star(P) = 1$ is trivial because that corresponds to zero
secrecy rate and the respective SRM solution is simply ${\bf W}=
{\bf 0}$.}.
\end{Theorem}

The remaining issue is to solve the equivalent SRM
problem~\eqref{Relx_SRM_MISOMEs}. Problem~\eqref{Relx_SRM_MISOMEs}
can be solved by using a bisection search methodology commonly used
in quasi-convex optimization. However, such a method would require
solving a sequence of (usually many) SDPs. Here we propose a more
efficient alternative where we exploit the problem structures and
recast~\eqref{Relx_SRM_MISOMEs} as an SDP. The idea is to use the
Charnes-Cooper transformation~\cite{Charnes1962,Hsin2008}.
%~\cite{Charnes1962,Chen2005,Hsin2008}.
Let us consider the following transformation of the transmit
covariance:
\[ {\bf W} = {\bf Z}/\xi \]
 for some ${\bf Z} \succeq {\bf 0}$, $\xi > 0$.
The change of variables above enables us to
re-express~\eqref{Relx_SRM_MISOMEs} as
\begin{subequations}\label{Relx_SRM_Transform}
\begin{align}
\displaystyle \min_{{\bf Z}, \xi } &  ~ \displaystyle
\max_{k=1,\ldots,K} \frac{ \xi + {\rm Tr}( {\bf G}_k {\bf G}_k^H {\bf Z}) }{ \xi + {\rm Tr}({\bf h}{\bf h}^H {\bf Z}) } \label{Relx_SRM_Transform_a}\\
{\rm s.t.} & ~ {\rm Tr}({\bf Z}) \leq \xi P, \label{Relx_SRM_Transform_b}\\
&  ~ {\bf Z} \succeq {\bf 0}, ~ \xi > 0,\label{Relx_SRM_Transform_c}
\end{align}
\end{subequations}
which may further be reformulated as an SDP
\begin{subequations}\label{Relx_SRM_Transform_SDP}
\begin{align}
\min_{ {\bf Z}, \xi, \tau } & ~ \tau \label{Relx_SRM_Transform_SDP_a}\\
{\rm s.t.} & ~ \xi + {\rm Tr}( {\bf G}_k {\bf G}_k^H {\bf Z}) \leq \tau, ~ k=1,\ldots,K, \label{Relx_SRM_Transform_SDP_b}\\
& ~ \xi + {\rm Tr}({\bf h}{\bf h}^H {\bf Z}) = 1, \label{Relx_SRM_Transform_SDP_c}\\
& {\rm Tr}({\bf Z}) \leq \xi P, \label{Relx_SRM_Transform_SDP_d}\\
& {\bf Z} \succeq {\bf 0}, ~ \xi \geq
0.\label{Relx_SRM_Transform_SDP_e}
\end{align}
\end{subequations}
Note
that~\eqref{Relx_SRM_Transform_SDP_a}-\eqref{Relx_SRM_Transform_SDP_b}
is a consequence of the standard epigraph reformulation (see the
literature, e.g.,~\cite{Boyd2004}), and the
constraint~\eqref{Relx_SRM_Transform_SDP_c} is additionally
introduced to fix the denominator of the objective function
in~\eqref{Relx_SRM_Transform_a} which is without loss of generality.
The relatively more crucial part with the reformulation lies in
replacing $\xi > 0$ in~\eqref{Relx_SRM_Transform_c} by $\xi \geq 0$
in~\eqref{Relx_SRM_Transform_SDP_e}. We show that this will not
cause a problem:

\begin{Prop}
The SDP~\eqref{Relx_SRM_Transform_SDP} is equivalent to Problem
~\eqref{Relx_SRM_Transform}. The former is also equivalent to the
SRM problem~\eqref{SRM_MISOMEs_reform} through the relation ${\bf W}
= {\bf Z}/\xi$.
\end{Prop}

{\it Proof:} \ The remaining, nontrivial part is when $\xi= 0$
in~\eqref{Relx_SRM_Transform_SDP_e}. Suppose that this is true.
Then, by~\eqref{Relx_SRM_Transform_SDP_d} and ${\bf Z} \succeq {\bf
0}$, we must have ${\bf Z}= {\bf 0}$. The
constraint~\eqref{Relx_SRM_Transform_SDP_c} is then violated as a
consequence. Thus, a feasible point
of~\eqref{Relx_SRM_Transform_SDP} must not have $\xi= 0$. This prove
that Problem~\eqref{Relx_SRM_Transform_SDP} is equivalent to
Problem~\eqref{Relx_SRM_Transform}. The solution equivalence of
Problems~\eqref{Relx_SRM_Transform} and \eqref{SRM_MISOMEs_reform}
through ${\bf W} = {\bf Z}/\xi$ follows from the discussion above
and Theorem 1.
\hfill $\blacksquare$ \\

Concluding, we have shown that the SRM
problem~\eqref{SRM_MISOMEs_reform} can essentially be solved by
using SDP; see Theorem 1 and Proposition 2. Moreover, Theorem 1
reveals that the SRM solution must be of rank one, which implies
that transmit beamforming is generally the optimal transmit strategy
for the SRM problem.

\section{Secrecy-Rate Optimization with Imperfect CSI}\label{Robust_SRM}
The MISO secrecy problems solved in the
previous sections have been based on a premise that the CSIs of Bob
and Eves are perfectly known to Alice. In this section, we extend
our results to the imperfect CSI case. We will consider robust
MISO secrecy-rate formulations that cater for CSI uncertainties in
the worst-case sense. The proposed robust secrecy-rate problems are
more complex in structures and more challenging to solve than their
perfect-CSI counterparts, but we will show that these problems can
still be turned to SDPs.
% by employing the approach established in the last section.
The robust formulations will be presented in the first subsection,
while the SDP solutions to the robust formulations will be described
in the second and third subsections.

%The MISO secrecy problems solved in the previous sections have been
%based on a premise that the CSIs of Bob and Eves are perfectly known
%to Alice. In this section, we extend our results to the imperfect
%CSI scenario. We will consider robust MISO secrecy-rate formulations
%that cater for CSI uncertainties. In the first three subsections, we
%focus on a deterministic uncertainty model and solve the robust
%secrecy-rate problem in the worst-case sense. The proposed
%worst-case secrecy-rate problems are more complex in structures and
%more challenging to solve than their perfect-CSI counterparts, but
%we will show that these problems can still be turned to SDPs. In the
%last subsection, we consider another kind of uncertainty
%model---stochastic model and formulate the robust secrecy-rate
%problem in the probabilistic sense. By employing the approach
%established in the worst-case subsections, we propose a conservative
%approach to solving the probabilistic constrained problem.

\subsection{Robust Secrecy-Rate Problem Formulations}\label{Robust_SRM_formulation}

Our model assumption for imperfect CSI is based on the deterministic
model~\cite{Shenouda2007,Vucic2009,Wang2009}. We model the
Alice-to-Bob channel and the Alice-to-Eve channels respectively by
\begin{align*}
{\bf h} & = \bar{\bf h} + \Delta{\bf h}, \\
{\bf G}_k & = \bar{\bf G}_k + \Delta{\bf G}_k, \quad k=1,\ldots,K.
\end{align*}
Here, $\bar{\bf h}$ and $\bar{\bf G}_k$ are the channel means and
they are known to Alice; $\Delta{\bf h}$ and $\Delta{\bf G}_k$
represent the channel uncertainties. These uncertainties are assumed
to be deterministic unknowns with bounds on their magnitudes:
\begin{align*}
\| \Delta{\bf h} \|_2 = \| {\bf h} - \bar{\bf h} \|_2 & \leq \varepsilon_b, \\
\| \Delta{\bf G}_k \|_F = \| {\bf G}_k - \bar{\bf G}_k \|_F & \leq
\varepsilon_{e,k}, \quad k=1,\ldots,K,
\end{align*}
for some $\varepsilon_b, \varepsilon_{e,1}, \ldots,
\varepsilon_{e,K} > 0$.

The proposed robust SRM formulation is as follows:
\begin{equation}\label{rob_SRM}
R^\star(P)= \max_{ \substack{ {\bf W} \succeq {\bf 0}, \\ {\rm
Tr}({\bf W}) \leq P } } \psi ({\bf W})
\end{equation}
where $\psi ({\bf W})$ is the worst-case secrecy rate function,
defined by
\begin{align}\label{rob_SRM_part}
\psi ({\bf W}) & = \min_{k=1,\ldots,K} \psi_k({\bf W}),\\
\psi_k ({\bf W}) & = \min_{ \substack{ {\bf h} \in \mathcal{B}_b, \\
{\bf G}_k \in \mathcal{B}_{e,k} } }
\log ( 1 + {\bf h}^H {\bf W} {\bf h} ) - \log \det( {\bf I} + {\bf G}_k^H {\bf W}{\bf G}_k), \\
\mathcal{B}_b & = \{ ~ {\bf h} \in \mathbb{C}^{N_t} ~|~ \| {\bf h}- \bar{\bf h} \|_2 \leq \varepsilon_b ~ \}, \\
\label{rob_SRM_part_end}
\mathcal{B}_{e,k} & = \{ ~ {\bf G}_k \in \mathbb{C}^{N_t \times
N_{e,k}} ~|~ \| {\bf G}_k - \bar{\bf G}_k \|_F \leq
\varepsilon_{e,k} ~ \}.
\end{align}
The function $\psi_k(\cdot)$ represents the worst secrecy-rate
function among all channel possibilities, for the $k$th Eve. The
resulting design~\eqref{rob_SRM} is a conservative one; from its
formulation we see that the secrecy rate will be guaranteed to be no
less than the worst-case optimum $R^\star(P)$ for any channel
possibilities (described by $\mathcal{B}_b$, $\mathcal{B}_{e,k}$).

Based on the same philosophy as in the last section, we can also formulate a
worst-case robust SRC design:
\begin{equation}\label{rob_SRC}
\begin{array}{rl}
\displaystyle P^\star(R) = \min_{ {\bf W} \succeq {\bf 0} } & {\rm Tr}({\bf W})     \\
{\rm s.t.} &  \displaystyle \psi ({\bf W}) \geq R,
\end{array}
\end{equation}
i.e., minimizing the transmit power while satisfying a secrecy rate
specification $R$ in the worst-case sense. This robust SRC problem
is interesting in its own rights, and, like the perfect-CSI solution
established in the last section, solving the robust SRC problem will
provide us with a crucial key to solving the robust SRM problem.

\subsection{The Robust Secrecy-Rate Constrained
Problem}\label{Robust_SRM_SRC}

We consider the robust SRC (R-SRC) problem~\eqref{rob_SRC} in this
subsection, where we aim to develop an SDP solution to R-SRC
optimization. From~\eqref{rob_SRM_part}-\eqref{rob_SRC},
Problem~\eqref{rob_SRC} can be expressed as
\begin{equation}\label{rob_SRC_reform}
\begin{array}{rl}
\displaystyle P^\star(R) = \min_{ {\bf W} \succeq {\bf 0} } & {\rm Tr}({\bf W})     \\
{\rm s.t.} &  \displaystyle 2^{-R} \geq \frac{ \displaystyle \max_{ {\bf G}_k \in \mathcal{B}_{e,k} } \det( {\bf I} + {\bf G}_k^H {\bf W} {\bf G}_k ) }{ \displaystyle \min_{ {\bf h} \in \mathcal{B}_b } 1 + {\bf h}^H {\bf W}{\bf h} }, \\
& \quad k=1,\ldots,K.
\end{array}
\end{equation}
Our approach to solving~\eqref{rob_SRC_reform} is somehow
reminiscent of its non-robust counterpart in Section~\ref{SRM_SRC}.
We use Lemma~1 to relax the constraints in~\eqref{rob_SRC_reform},
and obtain the following relaxed problem:
\begin{equation}\label{Relx_rob_SRC}
\begin{array}{rl}
\displaystyle P^\star(R) \geq P_{relax}^\star(R) = \min_{ {\bf W} \succeq {\bf 0} } & {\rm Tr}({\bf W})   \\
{\rm s.t.} &  \displaystyle 2^{-R} \geq \frac{ \displaystyle \max_{ {\bf G}_k \in \mathcal{B}_{e,k} } 1 + {\rm Tr}({\bf G}_k^H {\bf W} {\bf G}_k ) }{ \displaystyle \min_{ {\bf h} \in \mathcal{B}_b } 1 + {\bf h}^H {\bf W}{\bf h} },\\
& \quad k=1,\ldots,K.
\end{array}
\end{equation}
Our goals are then to turn~\eqref{Relx_rob_SRC} to a tractable
convex problem, and to show that the relaxation
in~\eqref{Relx_rob_SRC} is tight by proving the rank-one solution
structure of~\eqref{Relx_rob_SRC}.

In fact, the relaxed R-SRC problem~\eqref{Relx_rob_SRC} can be
recast as an SDP. To do so, the first step is to
reformulate~\eqref{Relx_rob_SRC} as
\begin{subequations}\label{Relx_rob_SRC_reform}
\begin{align}
\min_{ {\bf W} \succeq {\bf 0}, \theta } & ~ {\rm Tr}({\bf W}) \label{Relx_rob_SRC_reform_a}     \\
{\rm s.t.} & ~ \min_{ {\bf h} \in \mathcal{B}_b } 1 + {\bf h}^H {\bf W}{\bf h} \geq \theta, \label{Relx_rob_SRC_reform_b}\\
& ~ 2^{-R} \theta \geq \max_{ {\bf G}_k \in \mathcal{B}_{e,k} } 1 + {\rm Tr}({\bf G}_k^H {\bf W} {\bf G}_k ), \label{Relx_rob_SRC_reform_c}\\
& \quad \quad k=1,\ldots,K, \nonumber
\end{align}
\end{subequations}
where we have added a slack variable $\theta$ to decouple the
fractional functions in the original constraints
in~\eqref{Relx_rob_SRC}. Problem~\eqref{Relx_rob_SRC_reform} is
already a convex problem in principle, but with semi-infinite
constraints as seen in~\eqref{Relx_rob_SRC_reform_b}
and~\eqref{Relx_rob_SRC_reform_c}. To make the problem more
tractable to solve and analyze, the second step is to
turn~\eqref{Relx_rob_SRC_reform_b} and~\eqref{Relx_rob_SRC_reform_c}
to linear matrix inequalities (LMIs), using the
$\mathcal{S}$-procedure:

\begin{Lemma}[ {$\mathcal{S}$-procedure~\cite{Boyd2004}} ]
Let
\[ f_k({\bf x}) = {\bf x}^H {\bf A}_k {\bf x} + 2 {\rm Re}\{ {\bf b}_k^H {\bf x} \} + c_k \]
for $k=1,2$, where ${\bf A}_k \in \mathbb{H}^n$, ${\bf b}_k \in
\mathbb{C}^n$, $c_k \in \mathbb{R}$. The implication $f_1({\bf x})
\leq 0 \Rightarrow f_2({\bf x}) \leq 0$ holds if and only if there
exists $\mu \geq 0$ such that
\[  \mu \begin{bmatrix} {\bf A}_1 & {\bf b}_1 \\ {\bf b}_1^H & c_1 \end{bmatrix}
- \begin{bmatrix} {\bf A}_2 & {\bf b}_2 \\ {\bf b}_2^H & c_2
\end{bmatrix} \succeq {\bf 0}, \] provided that there exists a point
$\hat{\bf x}$ such that $f_1(\hat{\bf x}) < 0$.
\end{Lemma}

To apply the $\mathcal{S}$-procedure to the
constraint~\eqref{Relx_rob_SRC_reform_b}, we substitute the
representation ${\bf h} = \bar{\bf h} + \Delta{\bf h}$
into~\eqref{Relx_rob_SRC_reform_b} and
re-express~\eqref{Relx_rob_SRC_reform_b} as:
\begin{equation}\label{uncertainty_implication}
\Delta{\bf h}^H \Delta{\bf h} \leq \varepsilon_b^2 \Longrightarrow
\Delta{\bf h}^H {\bf W} \Delta{\bf h} + 2 {\rm Re}\{ \bar{\bf h}^H
{\bf W} \Delta{\bf h} \} + \bar{\bf h}^H {\bf W} \bar{\bf h} + 1 -
\theta \geq 0.
\end{equation}
Using the $\mathcal{S}$-procedure, we
transform~\eqref{uncertainty_implication} to an LMI
\begin{equation}\label{rob_SRC_LMI_1}
{\bf T}_b( {\bf W}, \lambda_b, \theta ) \triangleq
\begin{bmatrix}
\lambda_b {\bf I}_{N_t} + {\bf W} &   {\bf W}\bar{\bf h}   \\
\bar{\bf h}^H {\bf W} & -\lambda_b \varepsilon_b^2 - \theta +
\bar{\bf h}^H {\bf W} \bar{\bf h} + 1
\end{bmatrix} \succeq {\bf 0},
\end{equation}
for some $\lambda_b \geq 0$. Similarly,
\eqref{Relx_rob_SRC_reform_c} is equivalent to the following
implication:
\begin{equation}
\Delta\bm{g}_k^H \Delta\bm{g}_k \leq \varepsilon_{e,k}^2
\Longrightarrow \Delta\bm{g}_k^H \bm{\mathcal{W}}_k \Delta\bm{g}_k +
2 {\rm Re}\{\bar{\bm{g}}_k^H \bm{\mathcal{W}}_k \Delta\bm{g}_k \} +
\bar{\bm{g}}_k^H \bm{\mathcal{W}}_k \bar{\bm{g}}_k + 1 -2^{-R}
\theta \leq 0,
\end{equation} where
$\bm{\mathcal{W}}_k = {\bf I}_{N_{e,k}} \otimes {\bf W}$ and
$\bar{\bm{g}}_k = {\rm vec}(\bar{\bf G}_k)$. By the
$\mathcal{S}$-procedure, the above implication can be re-expressed
as the following LMI:
\begin{equation}\label{rob_SRC_LMI_2}
{\bf T}_{e,k}( {\bf W}, \lambda_{e,k}, \theta ) \triangleq
\begin{bmatrix}
\lambda_{e,k} {\bf I}_{N_{e,k}N_t} - \bm{\mathcal{W}}_k &   - \bm{\mathcal{W}}_k \bar{\bm{g}}_k   \\
-\bar{\bm{g}}_k^H \bm{\mathcal{W}}_k  & -\lambda_{e,k}
\varepsilon_{e,k}^2 - \bar{\bm{g}}_k^H \bm{\mathcal{W}}_k
\bar{\bm{g}}_k + 2^{-R}\theta - 1
\end{bmatrix} \succeq {\bf 0},
\end{equation}
for some $\lambda_{e,k} \geq 0$, $k=1,\ldots,K$.
Substituting~\eqref{rob_SRC_LMI_1} and~\eqref{rob_SRC_LMI_2} back
into~\eqref{Relx_rob_SRC_reform}, we obtain the following SDP
\begin{subequations}\label{Relx_rob_SRC_SDP}
\begin{align}
P^\star_{relax}(R)= \min_{
{\bf W}, \theta, \lambda_b, \bm{\lambda}_e }
 &  ~ {\rm Tr}({\bf W})   \label{Relx_rob_SRC_SDP_a} \\
{\rm s.t.}  &  ~ {\bf T}_b( {\bf W}, \lambda_b, \theta ) \succeq {\bf 0},   \label{Relx_rob_SRC_SDP_b}\\
&  ~  {\bf T}_{e,k}( {\bf W}, \lambda_{e,k}, \theta ) \succeq {\bf 0}, ~ k=1,\ldots,K,   \label{Relx_rob_SRC_SDP_c}\\
& ~ {\bf W} \succeq {\bf 0}, ~ \lambda_b \geq 0, ~
{\lambda}_{e,k} \geq 0,~k=1,\ldots,K . \label{Relx_rob_SRC_SDP_d}
\end{align}
\end{subequations}
%\begin{subequations}\label{Relx_rob_SRC_SDP}
%\begin{align}
%P^\star_{relax}(R)= \min_{ {\bf W}, \theta, \lambda_b, {\color{blue} \{ {\lambda}_{e,k} \}_{k=1}^{K} } }   &  ~ {\rm Tr}({\bf W})   \label{Relx_rob_SRC_SDP_a} \\
%{\rm s.t.}  &  ~ {\bf T}_b( {\bf W}, \lambda_b, \theta ) \succeq {\bf 0},   \label{Relx_rob_SRC_SDP_b}\\
%&  ~  {\bf T}_{e,k}( {\bf W}, \lambda_{e,k}, \theta ) \succeq {\bf 0}, ~ k=1,\ldots,K,   \label{Relx_rob_SRC_SDP_c}\\
%& ~ {\bf W} \succeq {\bf 0}, ~ \lambda_b \geq 0, ~ {\color{blue}
%{\lambda}_{e,k} \geq 0,~k=1,\ldots,K }. \label{Relx_rob_SRC_SDP_d}
%\end{align}
%\end{subequations}
where $\bm{\lambda}_e= [~ \lambda_{e,1},\ldots,\lambda_{e,K} ~]^T$.
The SDP~\eqref{Relx_rob_SRC_SDP}, as an equivalent form of the
relaxed R-SRC problem, can be solved conveniently by available SDP
solvers. In addition to this, the problem structures
of~\eqref{Relx_rob_SRC_SDP} enable us to analyze the solution
optimality of the relaxed R-SRC problem via the KKT conditions:
\begin{Prop}\label{prop:rob_PM_rank1}
Consider the relaxed R-SRC problem~\eqref{Relx_rob_SRC_SDP} for the
case of $R > 0$, Also, suppose that Problem~\eqref{Relx_rob_SRC_SDP}
is feasible.
%Suppose that the relaxed R-SRC problem~\eqref{Relx_rob_SRC_SDP} is
%{\color{blue}feasible for $R
%> 0$}.
Then, the optimal solution of~\eqref{Relx_rob_SRC_SDP} must be of rank one and
unique.
\end{Prop}

It is interesting to note that the end results of Proposition 3 are
exactly the same as those of its non-robust counterpart, Proposition
1. However, to prove Proposition 3 is considerably more complicated,
owing to the more complex LMIs in \eqref{Relx_rob_SRC_SDP_b} and
\eqref{Relx_rob_SRC_SDP_c}. The proof of Proposition 3 is given
in Appendix C.

We complete this subsection by using Proposition 3 to verify that
the relaxed R-SRC problem~\eqref{Relx_rob_SRC_SDP} is tight:
\begin{Corollary}
Consider the R-SRC problem~\eqref{rob_SRC_reform} for the case of $R > 0$,
and suppose that problem~\eqref{rob_SRC_reform} is feasible.
%Suppose that the R-SRC problem~\eqref{rob_SRC_reform} is
%{\color{blue}feasible for $R > 0$}.
The relaxed R-SRC
problem~\eqref{Relx_rob_SRC_SDP} exactly solves the R-SRC problem,
in the sense that the optimal solutions of~\eqref{rob_SRC_reform}
and~\eqref{Relx_rob_SRC_SDP} are equivalent to one another.
Moreover, the optimal R-SRC solution is unique and of rank one.
\end{Corollary}

{\it Proof:} \ The proof is essentially identical to that of
Corollary 1, and thus is omitted for brevity.
\hfill $\blacksquare$ \\

\subsection{The Robust Secrecy-Rate Maximization
Problem}\label{Robust_SRM_SRM}

With the R-SRC results established in the previous subsection, we
are now ready to develop an SDP solution to the robust SRM (R-SRM)
problem~\eqref{rob_SRM}. Problem~\eqref{rob_SRM} can be expressed as
\begin{equation}\label{rob_SRM_reform}
\gamma^\star(P) = \min_{ \substack{{\bf W} \succeq {\bf 0}, \\ {\rm
Tr}({\bf W}) \leq P } } \max_{k=1,\ldots,K} \frac{ \displaystyle
\max_{ {\bf G}_k \in \mathcal{B}_{e,k} } \det( {\bf I} + {\bf G}_k^H
{\bf W} {\bf G}_k ) }{ \displaystyle \min_{ {\bf h} \in
\mathcal{B}_b } 1 + {\bf h}^H {\bf W}{\bf h} }
\end{equation}
where $0 < \gamma^\star(P) \leq 1$. Applying Lemma~1
to~\eqref{rob_SRM_reform} yields the following relaxed R-SRM problem
\begin{equation}\label{Relx_rob_SRM}
\gamma^\star(P)  \geq \gamma_{relax}^\star(P) = \min_{
\substack{{\bf W} \succeq {\bf 0}, \\ {\rm Tr}({\bf W}) \leq P } }
\max_{k=1,\ldots,K} \frac{ \displaystyle \max_{ {\bf G}_k \in
\mathcal{B}_{e,k} } 1 + {\rm Tr}({\bf G}_k^H {\bf W} {\bf G}_k ) }{
\displaystyle \min_{ {\bf h} \in \mathcal{B}_b } 1 + {\bf h}^H {\bf
W}{\bf h} }.
\end{equation}
We again investigate two issues: the tightness of the relaxed R-SRM
problem, and the possibility of converting~\eqref{Relx_rob_SRM} to a
convex problem.

The relaxed R-SRM problem is tight. We show the following:
\begin{Theorem}
Consider the R-SRM problem~\eqref{rob_SRM_reform} for the case of
$0< \gamma^\star(P) < 1$. The relaxed R-SRM
problem~\eqref{Relx_rob_SRM} exactly solves the R-SRM problem, in
the sense that the optimal solutions of~\eqref{rob_SRM_reform}
and~\eqref{Relx_rob_SRM} are equivalent to one another. Moreover,
the optimal R-SRM solution is unique and of rank one.
\end{Theorem}

{\it Proof:} \ The proof of Theorem 2 is essentially the same as
that of its non-robust counterpart, Theorem~1, and here we provide
only the outline. The idea is to consider the relaxed R-SRC problem
\begin{equation}\label{Relx_rob_SRC_construct}
\begin{array}{rl}
\displaystyle  \min_{ {\bf W} \succeq {\bf 0} } & {\rm Tr}({\bf W})     \\
{\rm s.t.} &  \displaystyle \gamma_{relax}^\star(P) \geq \frac{ \displaystyle \max_{ {\bf G}_k \in \mathcal{B}_{e,k} } 1 + {\rm Tr}({\bf G}_k^H {\bf W} {\bf G}_k ) }{ \displaystyle \min_{ {\bf h} \in \mathcal{B}_b } 1 + {\bf h}^H {\bf W}{\bf h} }, \\
& \quad k=1,\ldots,K.
\end{array}
\end{equation}
It is shown that the optimal solutions of~\eqref{Relx_rob_SRM}
and~\eqref{Relx_rob_SRC_construct} are equivalent to one another, by
following the same procedure as in Appendix B. The solution
equivalence of ~\eqref{Relx_rob_SRM}
and~\eqref{Relx_rob_SRC_construct}, together with Proposition 3,
enable us to deduce that~\eqref{Relx_rob_SRM} has a unique rank-one
solution. In the same spirit as the proof of Corollary~1, the unique
rank-one solution characteristic of~\eqref{Relx_rob_SRM} further
implies that $\gamma^\star(P) = \gamma_{relax}^\star(P)$ (recall
Lemma~1), and that the solution of~\eqref{Relx_rob_SRM} has to be
the solution of~\eqref{rob_SRM_reform}, and vice versa.
\hfill $\blacksquare$ \\

The relaxed (actually, equivalent) R-SRM
problem~\eqref{Relx_rob_SRM} can be transformed to an SDP. By
employing the Charnes-Cooper transformation and
$\mathcal{S}$-procedure, and by some careful manipulations, we show
that

\begin{Prop} \label{prop:R-SRM_reformulation}
Problem~\eqref{Relx_rob_SRM} is equivalent to the following SDP
\begin{subequations}\label{Relx_rob_SRM_SDP}
\begin{align}
\min_{ {\bf Z}, \xi,  \tau, \lambda_b, \bm{\lambda}_e } & ~ \tau \\
{\rm s.t.} & ~ {\bf M}_b( {\bf Z}, \lambda_b, \xi ) \succeq {\bf 0}, \\
& ~ {\bf M}_{e,k}( {\bf Z}, \lambda_{e,k}, \xi, \tau ) \succeq {\bf 0}, ~ k=1,\ldots,K, \\
& ~ {\rm Tr}( {\bf Z} ) \leq \xi P, \\
& ~ {\bf Z} \succeq {\bf 0}, ~ \xi \geq 0, ~ \lambda_b \geq 0, ~
\lambda_{e,k} \geq 0,~k=1,\ldots,K ,
\end{align}
\end{subequations}
%\begin{subequations}\label{Relx_rob_SRM_SDP}
%\begin{align}
%\min_{ {\bf Z}, \xi,  \tau, \lambda_b, {\color{blue} \{ \lambda_{e,k} \}_{k=1}^{K} }} & ~ \tau \\
%{\rm s.t.} & ~ {\bf M}_b( {\bf Z}, \lambda_b, \xi ) \succeq {\bf 0}, \\
%& ~ {\bf M}_{e,k}( {\bf Z}, \lambda_{e,k}, \xi, \tau ) \succeq {\bf 0}, ~ k=1,\ldots,K, \\
%& ~ {\rm Tr}( {\bf Z} ) \leq \xi P, \\
%& ~ {\bf Z} \succeq {\bf 0}, ~ \xi \geq 0, ~ \lambda_b \geq 0, ~
%{\color{blue}  \lambda_{e,k} \geq 0,~k=1,\ldots,K }.
%\end{align}
%\end{subequations}
where $\bm{\lambda}_e= [~ \lambda_{e,1},\ldots,\lambda_{e,K} ~]^T$,
\begin{equation*}
{\bf M}_b( {\bf Z}, \lambda_b, \xi ) \triangleq
\begin{bmatrix} \lambda_b {\bf I}_{N_t} + {\bf Z}  &  {\bf Z}\bar{\bf h} \\
\bar{\bf h}^H {\bf Z}  &  \bar{\bf h}^H {\bf Z} \bar{\bf h} + \xi -
\lambda_b \varepsilon^2_b -1
\end{bmatrix},
\end{equation*}
\begin{equation*}
{\bf M}_{e,k}( {\bf Z}, \lambda_{e,k}, \xi, \tau ) \triangleq
\begin{bmatrix} \lambda_{e,k} {\bf I}_{N_{e,k} N_t} - \bm{\mathcal{Z}}_k &  - \bm{\mathcal{Z}}_k \bar{\bm{g}}_k \\
- \bar{\bm{g}}_k^H \bm{\mathcal{Z}}_k  &  - \lambda_{e,k}
\varepsilon^2_{e,k} - \xi + \tau - \bar{\bm{g}}_k^H
\bm{\mathcal{Z}}_k \bar{\bm{g}}_k
\end{bmatrix},
\end{equation*}
and $\bm{\mathcal{Z}}_k = {\bf I}_{N_{e,k}} \otimes {\bf Z}$.
Specifically, Problems~\eqref{Relx_rob_SRM}
and~\eqref{Relx_rob_SRM_SDP} are equivalent through the relation
${\bf W}= {\bf Z}/\xi$.
\end{Prop}

We delegate the proof of Proposition~\ref{prop:R-SRM_reformulation} to Appendix~\ref{appendix:prop4},
since the key ideas behind the proof, the Charnes-Cooper transformation and $\mathcal{S}$-procedure, have been demonstrated in the preceding development.

In summary, we have shown that the nonconvex R-SRM problem
\eqref{rob_SRM_reform} can be equivalently solved by solving the
convex SDP \eqref{Relx_rob_SRM_SDP}; see Theorem 2 and Proposition
4.

%Recall that in the perfect CSI scenario studied in the previous
%section, transmit beamforming is found to be the optimal transmit
%strategy for the SRC and SRM designs in general. Interestingly, as
%indicated in Corollary~2 and Theorem~2, it remains valid that
%transmit beamforming is generally the optimal transmit strategy for
%the worst-case robust SRC and SRM designs.

Recall that in the perfect CSI scenario studied in the previous
section, transmit beamforming is shown to be the optimal transmit
strategy for the SRC and SRM designs in general.
This physical result remains valid for the worst-case robust SRC and SRM designs,
as indicated in Corollary~2 and Theorem~2.

%-----------------------------------------------------------------------------
\section{Simulation Results}\label{simulation}
%In this section, we provide several examples to verify that our
%proposed SDP solutions achieve better performances than other
%existing methods under both perfect and imperfect CSI scenarios. We
%assume all elements of ${\bf h}$ and ${\bf G}_k$, for all $k$ are
%independent and identically distributed (i.i.d.) complex Gaussian
%variables with mean zero and unit variance, and all results were
%averaged over 1000 randomly generated channel realizations.

We provide simulation results to illustrate the secrecy-rate performance gains of the proposed SDP solutions compared to some other existing methods.
We will first consider the perfect CSI case in the first subsection, and then the imperfect CSI case in the second subsection.

\subsection{The Perfect CSI Case}

The results to be presented in this subsection are based on the
following simulation settings, unless specified: At Alice, the
number of transmit antennas is $N_t= 10$, and the average transmit
power limit is $P= 3$\,dB. Three Eves are present ($K= 3$), and
their numbers of receive antennas are $N_{e,1}= \ldots = N_{e,K} =
3$. Perfect CSIs are assumed. At each trial of the simulations,
Bob's channel ${\bf h}$ is randomly generated following an
independent and identically distributed (i.i.d.) complex Gaussian
distribution with zero mean and unit variance. Similarly, each Eve's
channel ${\bf G}_k$ is randomly generated following an i.i.d.
complex Gaussian distribution with zero mean and variance
$\rho^2_e$.
%Note that $\rho^2_e$ describes the average received
%signal strength of an Eve relative to that of Bob--- $\rho_e > 1$
%means that every Eve has a stronger received signal strength than
%Bob, while $\rho_e < 1$ means the vice versa.
We fix $\rho^2_e= 1$, if not mentioned.
%(identical average signal strengths for Eves and Bob).
The simulation results were obtained based on an average of 1000 independent trials.

%The simulations evaluate the secrecy rates (see \eqref{SRM_MISOMEs}-\eqref{mutual_inf_diff} for the definition) of the following transmit designs:
The following transmit designs are tested in our simulations: the
proposed SDP solution \eqref{Relx_SRM_Transform_SDP} to solving the
SRM problem  \eqref{SRM_MISOMEs}, the projected-MRT method described
in Section~\ref{background_MultiEve}, and a simple method called
{\it plain-MRT} here. In plain-MRT we choose Bob's channel
direction, ${\bf h}/ \| {\bf h} \|$, as the transmit weight.
Specifically, plain-MRT sets ${\bf W}= (P/\| {\bf h}\|^2 ) {\bf h}
{\bf h}^H$ if the corresponding secrecy rate is positive; and ${\bf
W}= {\bf 0}$ otherwise. Plain-MRT is a suboptimal, arguably weak,
method since it ignores the presence of Eves. It is nonetheless
interesting to examine its secrecy-rate performance relative to SDP.

\begin{figure}[h]
\begin{center}
\subfigure[][]{\resizebox{.4\textwidth}{!}{\includegraphics{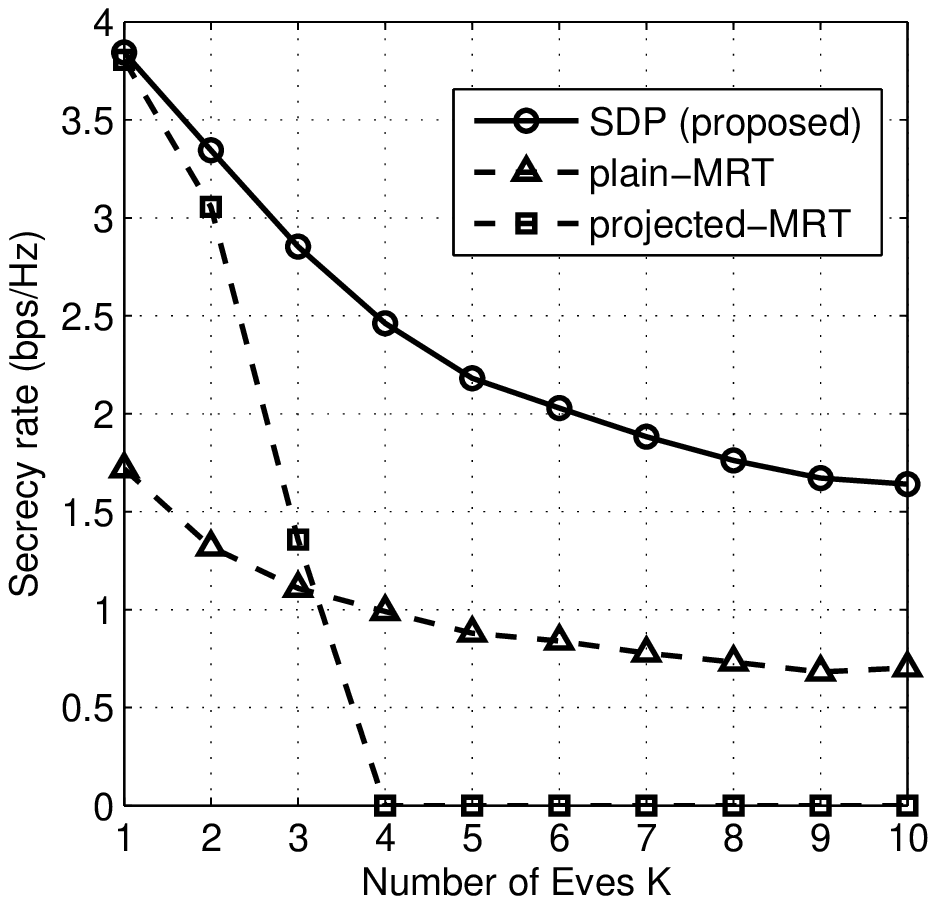}}}
\subfigure[][]{\resizebox{.4\textwidth}{!}{\includegraphics{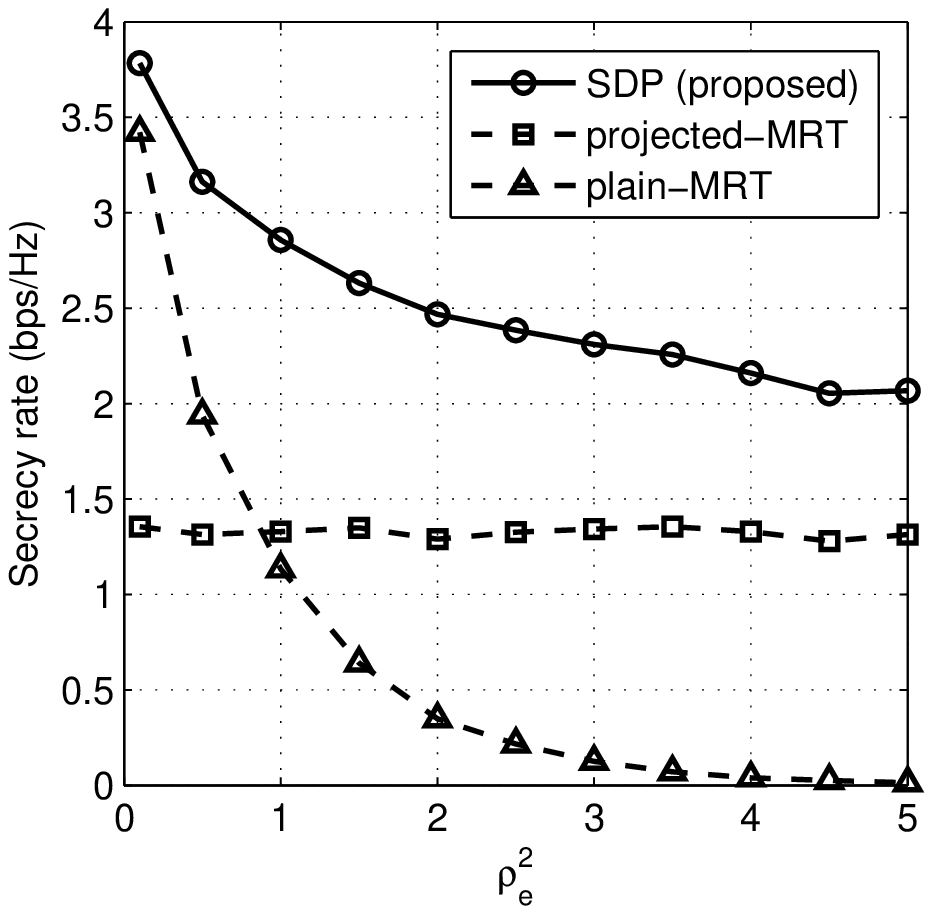}}}
\end{center}
\caption{Secrecy rates of the various methods versus (a) the number of Eves, and (b) the average channel strength of Eves.}
\label{fig:example1}
\end{figure}

\subsubsection{Secrecy rates versus the number of Eves}
Fig.~\ref{fig:example1}(a) shows the secrecy rate behaviors of the various methods when
we increase the number of Eves $K$.
We can see that the proposed SDP method yields better performance than the two other methods over the whole range of $K$ tested.
The secrecy rate of projected-MRT is able to approach that of SDP for $K \leq 2$, but becomes zero for $K > 3$;
the latter case is when the degree of freedom of all Eves combined, $\sum_{k=1}^K N_{e,k} = 3 K$, is higher than the transmit degree of freedom.
By contrast, the SDP method is able to provide a secrecy rate of higher than $1.5$bps/Hz even with $K= 10$.

\subsubsection{Secrecy rates versus Eves' received signal strength}
We investigate the impact of `near-far' effects on the secrecy rate behaviors;
i.e.,
how the various methods perform when Eves' received signal strength, characterized by $\rho_e^2$, changes.
Fig.~\ref{fig:example1}(b) shows the secrecy rates of the various methods with respect to $\rho_e^2$.
Note that $\rho_e > 1$
means that every Eve has a stronger received signal strength than
Bob, while $\rho_e < 1$ means the vice versa.
The following two phenomenons are observed for the MRT methods:
First, plain-MRT approaches SDP for small $\rho_e^2$, which makes sense since one may simply ignore weak Eves in the transmit design.
Second, projected-MRT has its secrecy rate invariant to $\rho_e^2$, which is due to its nulling process.
This means that projected-MRT can cope with strong Eves rather effectively.
Fig.~\ref{fig:example1}(b) also illustrates that the proposed SDP methods provide better performance than the two MRT methods, especially for $\rho_e^2 \geq 0.5$.

%ken:17_06_2010
%%------------------------------
%\subsubsection{Secrecy rate versus Eve's intercepted
%signal strength}
%
%In this example we investigate the `near-far effects' on the secrecy
%rate, that is, how the secrecy rates of the various methods are
%affected by Eve's channel strength $\rho_e^2$. From
%Fig.~\ref{fig:example2}, we can see that there are notable secrecy
%rate gaps between SDP and the other methods when $\rho_e^2 \ge 0.5$.
%Moreover, plain-MRT is very sensitive to the increase of $\rho_e^2$;
%its secrecy rate drops rapidly and approaches zero as $\rho_e^2$
%increases. In comparison, projected-MRT is almost invariant over all
%$\rho_e^2$'s.
%%-------------Fig near-far effect----------------
%\begin{figure}[H]
%\centering {
%\includegraphics[width=0.5\textwidth]{./Fig/NearFar_N10M3K3.eps}
%}\caption{Secrecy rate versus Eve's channel strength ($K=3$,
%$N_t=10$, $N_{e,k}=3$ for all $k$, $P$=3\,dB).}\label{fig:example2}
%\end{figure}

\begin{figure}[ht]
\centering {
\includegraphics[width=0.4\textwidth]{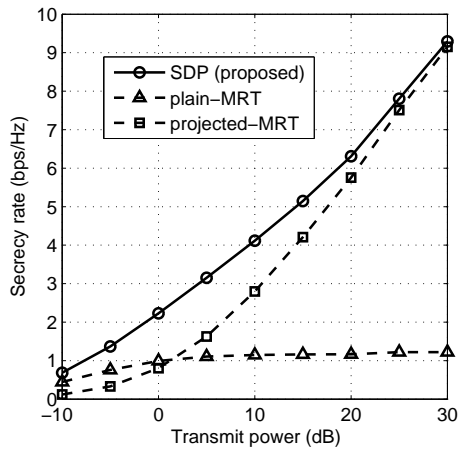}
}\caption{Secrecy rate versus transmit power.
%($N_t=10$, $K=3$, $N_{e,k}=3$, for all $k$).
}\label{fig:example3}
\end{figure}

\subsubsection{Secrecy rates versus the transmit power}
We are interested in evaluating the secrecy rate performance of the various methods with respect to the transmit power budget $P$.
The results are displayed in Fig.~\ref{fig:example3}.
Interestingly, we see that the secrecy rate of SDP appears to be approached by that of plain-MRT for small $P$, and by that of projected-MRT for large $P$.
\subsection{The Imperfect CSI Case}

The simulation settings in the imperfect CSI case are generally identical to those of the perfect CSI case above, except that the average power limit is increased to $P= 20$dB.
Regarding the imperfect CSI effects,
we define the following channel uncertainty ratios:
\begin{align*}
\alpha_{e,k} & = \frac{ \varepsilon_{e,k} }{ \sqrt{ {\rm E}\{ \| \bar{\bf G}_k \|_F^2 \} } }, \quad k=1,\ldots,K \\
\alpha_{b} & = \frac{ \varepsilon_b }{ \sqrt{ {\rm E}\{ \| \bar{\bf
h} \|^2 \} } },
\end{align*}
and use them to control the amount of channel uncertainties in the
simulations. We fix $\alpha_{e,1} = \ldots = \alpha_{e,K} \triangleq
\alpha_{e}$. We will choose $\alpha_b=0.03$, $\alpha_{e} =0.1$,
unless specified. The performance measure is the worst-case secrecy
rate defined in \eqref{rob_SRM}-\eqref{rob_SRM_part_end}. This
performance measure does not have a closed form, but can be computed
via SDP; the details are described in
Appendix~\ref{appendix:cal_worst_sec}.

We evaluate the performance of the robust SDP method developed in Section~\ref{Robust_SRM_SRM},
the non-robust SDP method (in Section~\ref{SRM_SRM}),
projected-MRT, and plain-MRT.
The latter three methods use the presumed CSIs $\bar{\bf h}, \bar{\bf G}_1, \ldots, \bar{\bf G}_K$ to perform transmit designs in the simulations, and then we evaluate the resultant worst-case secrecy rates.

%ken:17_06_2010
%Previous simulations are conducted under the perfect CSI assumption.
%In what follows, we will take the channel uncertainties into
%consideration and evaluate the worst-case secrecy rate of the robust
%and non-robust methods. Recall in
%Section~\ref{Robust_SRM_formulation} that the worst-case secrecy
%rate of a given transmit covariance ${\bf W}$ is
%$\displaystyle\min_{k=1,\ldots,K} g_k({\bf W})$, where $g_k({\bf
%W})$ is given in \eqref{rob_SRM_part}. This worst-case secrecy rate
%does not have a closed form, but can be numerically evaluated by
%using SDP methodology; the details are described in
%Appendix~\ref{appendix:cal_worst_sec}. In the following simulations,
%$\mathbf{\bar{G}}_k$ and $\mathbf{\bar{h}}$ are i.i.d. complex
%Gaussian with mean zero and unit variance and all results were
%averaged over 1000 randomly generated channel realizations.

\begin{figure}[htp]
\begin{center}
\subfigure[][]{\resizebox{.4\textwidth}{!}{\includegraphics{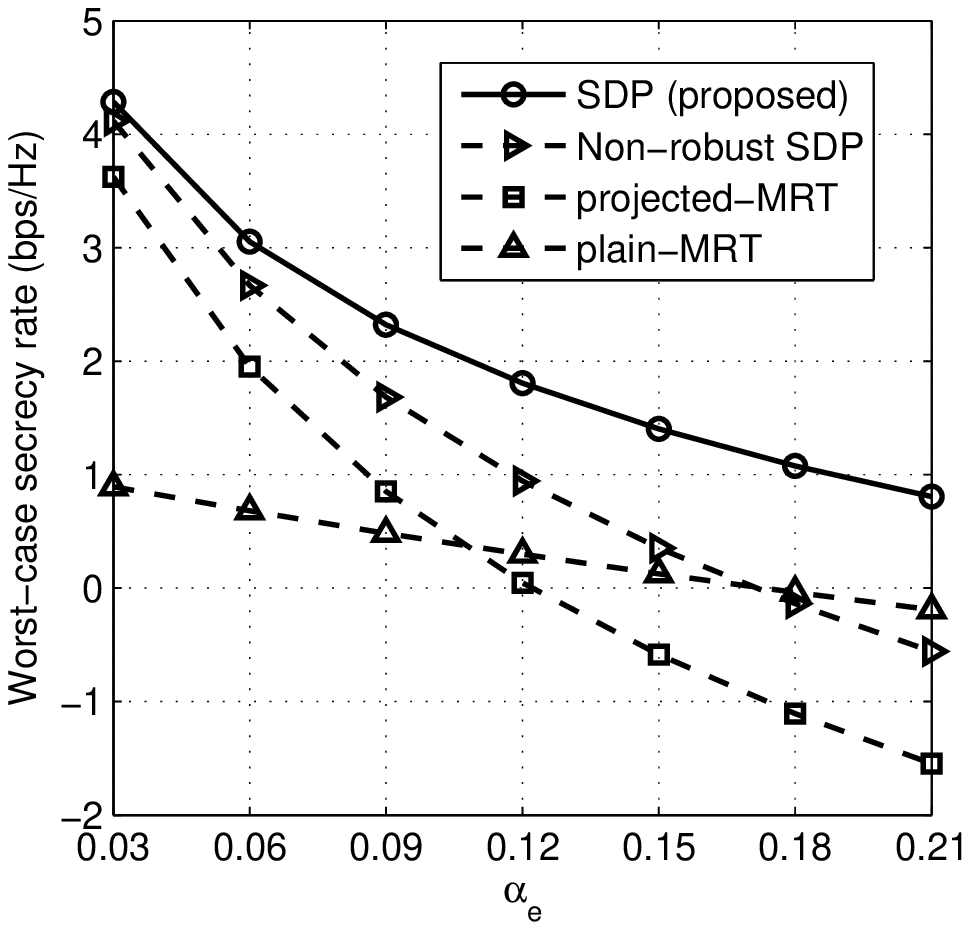}}}
\subfigure[][]{\resizebox{.4\textwidth}{!}{\includegraphics{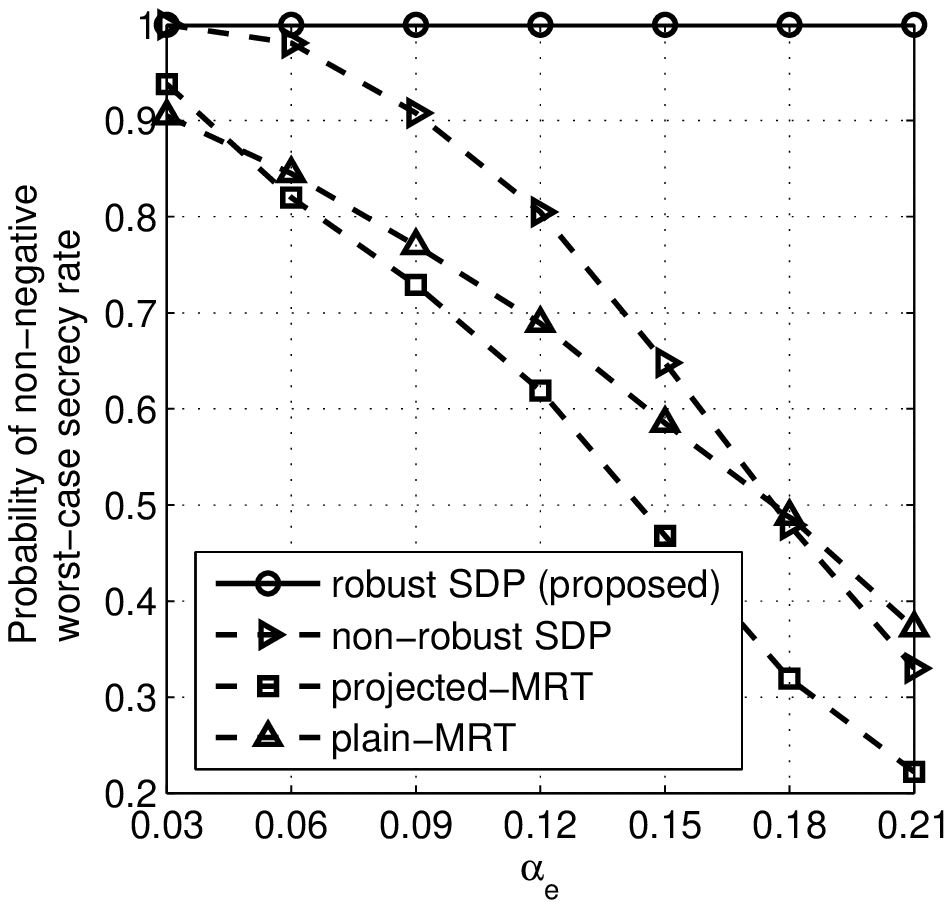}}}
\end{center}
\caption{ Secrecy rate performance with imperfect CSI. (a)
Worst-case secrecy rate versus Eves' channel uncertainty ratio; (b)
Probability of non-negative worst-case secrecy rate versus Eves'
channel uncertainty ratio. } \label{fig:example5}
\end{figure}

\subsubsection{Secrecy rate performance versus Eve's channel uncertainty ratio}
Fig.~\ref{fig:example5}(a) presents the worst-case secrecy rates of
the various methods versus Eve's channel uncertainty ratio
$\alpha_e$. As seen in the figure, the robust SDP method yields the
best worst-case secrecy rate among all the methods, especially when
$\alpha_e$ is large. Moreover, we observe a peculiar behavior---
that non-robust SDP, projected-MRT, and plain-MRT yield negative
worst-case secrecy rates for
%sufficiently large $\alpha_e$, say,
$\alpha_e > 0.18$. That is because these perfect-CSI-based methods
aim to provide non-negative secrecy rate results for the presumed
CSIs, but not for the actual CSIs. To get a better idea of how
sensitive the non-robust methods can be in the presence of imperfect
CSIs, in Fig.~\ref{fig:example5}(b) we show the probability of
non-negative secrecy rate; i.e., the chance that a method gives
non-negative worst-case secrecy rate under the 1000 independent
trials. One can see that the robust method always guarantees a
non-negative secrecy rate (which is expected from its design
formulation), and that the non-robust methods violate non-negative
secrecy rate quite seriously especially for large $\alpha_e$.

\begin{figure}[htp]
\begin{center}
\subfigure[][]{\resizebox{.40\textwidth}{!}{\includegraphics{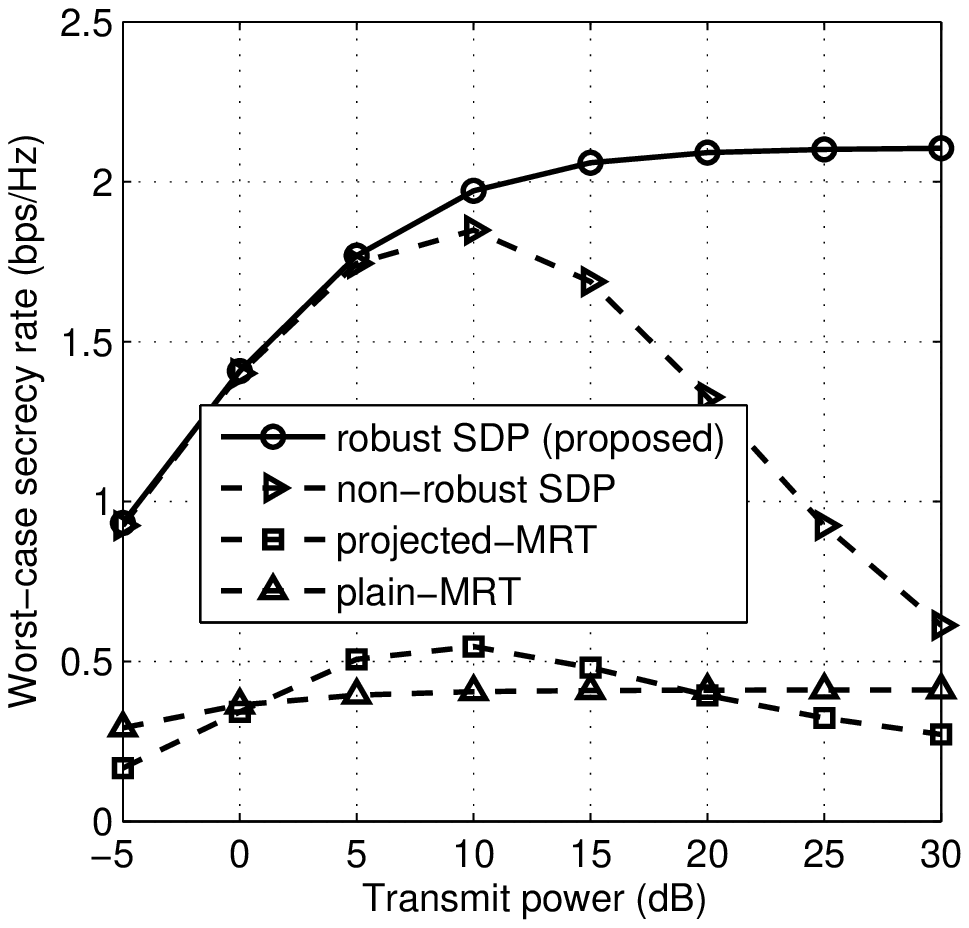}}}
\subfigure[][]{\resizebox{.40\textwidth}{!}{\includegraphics{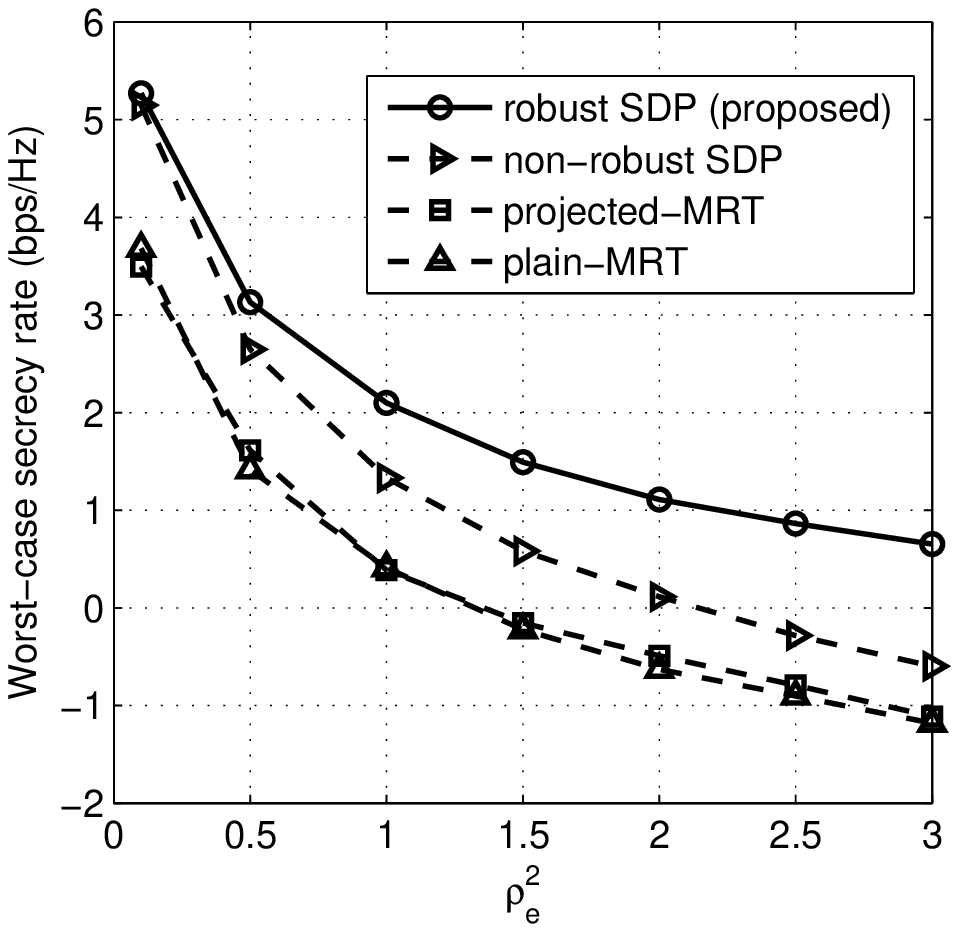}}}
\end{center}
\caption{
Secrecy rate performance with imperfect CSI.
(a) Worst-case secrecy rate versus transmit power;
(b) Worst-case secrecy rate versus Eves' received signal strength.
}
\label{fig:example4}
\end{figure}

\subsubsection{Worst-case secrecy rates versus transmit power, and Eves' received signal strength}
More results are shown to demonstrate the robustness of the proposed robust SDP method.
Fig.~\ref{fig:example4}(a) plots the worst-case secrecy rates of the various methods against the transmit power $P$.
As seen, the worst-case secrecy rate performance of the robust SDP method is better than those of the other methods.
Moreover, we observe that for non-robust SDP and projected-MRT,
keeping on increasing $P$ ends up decreasing the worst-case secrecy rate.
That is because increasing the power may also help improve eavesdroppers' receptions, if the transmit design does not take channel uncertainties into account.
Fig.~\ref{fig:example4}(b) plots the worst-case secrecy rates against $\rho_e^2$.
Again, the robust SDP method is seen to yield better worst-case secrecy rates than the other methods.

\section{Conclusion and Discussion}\label{conclusion}
To conclude, this paper has addressed the transmit covariance design
problem of
%maximizing the MISO secrecy rate overheard by multiple
%multi-antenna eavesdroppers,
MISO secrecy-rate maximization in the presence of multiple
multi-antenna eavesdroppers, using an effective SDP approach. Both
perfect and imperfect CSI cases are considered in our designs.
%We
%have shown a physically useful property that
We have shown by analysis that
transmit beamforming is generally the secrecy-rate optimal strategy %for the scenario under consideration.
 for the considered scenarios.
As illustrated by the simulations, the proposed SDP solutions
outperform some other existing methods.

It is worthwhile to mention some possible extensions of this work.
In~\cite{QLI_ispacs}, we have illustrated that the same SDP approach
can essentially be used to solve MISO SRM with more complex
covariance constraints, such as per-antenna power constraints, and
interference temperature constraints arising from CR.
In~\cite{QLI_icc}, we have considered a multicast extension; i.e.,
multiple Bobs receiving the same information. There, our SDP
approach is provably optimal only for some special cases such as the
$1$ Eve, $2$-to-$3$ Bobs setting; it serves as a tractable
approximation otherwise. Despite this, the SDP approach was
empirically shown to provide good lower and upper bounds for the
multicast secrecy rate.

%{\color{blue}Moreover, it is
%not difficult to see that the approach developed above can be easily
%extended to deal with multiple legitimate receivers~\cite{QLI_icc},
%and/or more complex covariance constraints, e.g., per-antenna power
%constraints and interference temperature constraints in cognitive
%radio systems~\cite{QLI_ispacs}. }

%In addition,
%one interesting extension to this paper is to take the artificial
%noise into the SRM design (the usefulness of the artificial noise in
%degrading Eves has been demonstrated in
%\cite{Jorswieck,Negi2005,Mukherjee09}). In this case, we need to
%jointly optimize the transmit covariance ${\bf W}$ and the
%artificial noise, which will result in a more complex and nonconvex
%problem.

Before closing this paper, we should highlight that some concurrent
research studies, e.g.,
\cite{Jorswieck,Negi2005,Mukherjee09,Liao11}, have demonstrated that
adding artificial noise (AN) in the transmit design is quite
effective in degrading Eves' receptions. Hence, a meaningful future
direction would be to extend this work to the AN-aided case,
optimizing the transmit design and AN jointly for the maximum
secrecy rate. The resultant design optimization is expected to be
more difficult than those tackled here, and it would be interesting
to see how
%the results in this paper
the proposed SDP approach may be used to help overcome the arising
design challenges. Moreover, while this paper has its theme on
optimal SRM designs, a parallel direction that is also meaningful is
to further investigate simple suboptimal SRM designs and their
aspects. For example, one intuitive design is to treat all the Eves
as one ``super-Eve'', and then employ the one-Eve optimal design
\eqref{solution_MISOME} which has a closed form. Owing to page
limit, such a possibility is not considered in this paper, but would
be another interesting future direction.

\section{Acknowledgment}
The authors would like to sincerely thank the anonymous reviewers
for their helpful and insightful comments.

%---------------------------------------------------------------------------
\appendix

\subsection{Proof of Proposition~\ref{prop:PM_rank1}} \label{appendix:prop1}

For Problem~\eqref{RPM_SRC_SDP}, let us first write out its
Lagrangian function:
\[ \mathcal{L} \big( {\bf W}, {\bf Y}, \bm{\mu} \big) = {\rm Tr}({\bf W}) -{\rm Tr}({\bf W} {\bf Y}) + \sum_{k=1}^{K}
\mu_k \left ( 2^R (1+ {\rm Tr}( {\bf G}_k {\bf G}_k^H {\bf W})) -
{\rm Tr}({\bf h} {\bf h}^H {\bf W}) -1 \right),\]
%\[ \mathcal{L} \big( {\bf W}, {\bf Y}, {\color{blue} \{ \mu_k \}_{k=1}^{K} } \big) = {\rm Tr}({\bf W}) -{\rm Tr}({\bf W} {\bf Y}) + \sum_{k=1}^{K}
%\mu_k \left ( 2^R (1+ {\rm Tr}( {\bf G}_k {\bf G}_k^H {\bf W})) -
%{\rm Tr}({\bf h} {\bf h}^H {\bf W}) -1 \right),\]
where $\bm{\mu} = [~ \mu_1,\ldots,\mu_K ~]^T$, $\mu_1,\ldots,\mu_K
\geq 0$ are the Lagrangian dual variable for the minimum
secrecy-rate constraints, and ${\bf Y} \in \mathbb{H}_{+}^{N_t}$ is
the Lagrangian dual variables for the constraint ${\bf W} \succeq
{\bf 0}$.
% {\color{blue} {\it (removed? Since Problem~\eqref{RPM_SRC_SDP} is
%feasible for $R>0$ and the constraints are linear, it is easy to see that the Slater
%condition~\cite{Boyd2004} is satisfied.)}} {\color{red} \it (there is something more than that!)}
%{\color{blue} Assuming Slater's condition~\cite{Boyd2004} as the constraint qualification,
%we show from  $\mathcal{L}\big(
%{\bf W}, {\bf Y}, \bm{\mu} \big)$
%above that the KKT conditions for a primal-dual point $\big({\bf W},{\bf
%Y}, \bm{\mu} \big)$ to be optimal is}
The corresponding KKT conditions are shown to be
\begin{subequations} \label{proof_p1_0}
\begin{align}
{\bf Y} & = \textstyle {\bf I} + 2^R \sum_{k=1}^K \mu_k {\bf G}_k
{\bf G}_k^H - ( \sum_{k=1}^K \mu_k ) {\bf h} {\bf h}^H,
\label{eq:proof_p1} \\
{\bf Y}{\bf W} & = {\bf 0},
\label{eq:proof_p1_2} \\
& 1+ {\rm Tr}({\bf h}{\bf h}^H {\bf W} ) \geq 2^R  (1 + {\rm Tr}({\bf G}_k {\bf G}_k^H {\bf W}  ) ), ~~ \forall k \label{eq:proof_p1_3}\\
{\bf W} & \succeq {\bf 0}, \quad {\bf Y} \succeq {\bf 0}, \quad
\mu_k \geq 0, ~~ k=1,\ldots,K.
\end{align}
\end{subequations}
Note that in general, problem~\eqref{RPM_SRC_SDP} satisfies Slater's constraint qualification condition:
If problem~\eqref{RPM_SRC_SDP} has a feasible point, then one can prove, by construction, that there exists a strictly feasible point for problem~\eqref{RPM_SRC_SDP}.
As a result, strong duality holds and
the KKT conditions are the necessary conditions for a primal-dual point
$\big({\bf W},{\bf Y}, \bm{\mu} \big)$ to be optimal.

The key to showing the rank-one structure of ${\bf W}$ lies in
\eqref{eq:proof_p1}. Let
\[ {\bf B} = \textstyle {\bf I} + 2^R \sum_{k=1}^K \mu_k {\bf G}_k {\bf G}_k^H. \]
We see that ${\bf B}$ is positive definite, and thus has full rank.
By letting $\rho= \sum_{k=1}^K \mu_k \geq 0$ and by denoting ${\bf
B}^{1/2}$ as a positive definite square root of ${\bf B}$, we have
that
\begin{align*}
{\rm rank}({\bf Y})
%& = \textstyle {\rm rank}( {\bf B}^{1/2} ( {\bf I} - \rho ( {\bf B}^{-1/2}{\bf h} ) ({\bf B}^{-1/2}{\bf h})^H ) {\bf B}^{1/2} ) \\
& \equiv {\rm rank}( {\bf B}^{-1/2} {\bf Y} {\bf B}^{-1/2} )  \\
& = {\rm rank}(  {\bf I} - \rho ( {\bf B}^{-1/2}{\bf h} ) ({\bf
B}^{-1/2}{\bf h})^H ) \geq N_t - 1,
\end{align*}
i.e., ${\rm rank}({\bf Y})$ is either $N_t$ or $N_t -1$. For ${\rm
rank}({\bf Y})= N_t$, \eqref{eq:proof_p1_2} can only be satisfied by
${\bf W} = {\bf 0}$. However, ${\bf W}= {\bf 0}$ violates
\eqref{eq:proof_p1_3} when $R>0$. For ${\rm rank}({\bf Y})= N_t-1$,
\eqref{eq:proof_p1_2} is achieved only when ${\bf W}$ lies in the
nullspace of ${\bf Y}$, the dimension of which is one. This means
that any optimal ${\bf W}$ must be of rank one.

The proof above has shown that any primal optimal solution ${\bf W}$
of \eqref{RPM_SRC_SDP} must be of rank one for $R>0$. Next, we
consider the uniqueness of the optimal ${\bf W}$. Suppose that there
are two distinct optimal solutions, say ${\bf W}_{1}$ and ${\bf
W}_{2}$, which satisfy ${\rm rank}({\bf W}_{1}) = {\rm rank}({\bf W}_{2})=1$.
{
It can be easily shown that
the subspaces spanned by
${\bf W}_{1}$ and ${\bf W}_{2}$ must be different in order for ${\bf W}_{1}$ and ${\bf
W}_{2}$ to be distinct; i.e.,
$\mathcal{R}({\bf W}_{1}) \neq \mathcal{R}({\bf W}_{2})$
where $\mathcal{R}( \cdot )$ denotes the range space of the argument.
}
%
%Then we have rank(${\bf W}_{1}$)=rank(${\bf W}_{2}$)=1.
%Moreover, it is easy to verify that ${\bf W}_1 \ne \alpha {\bf W}_2$
%for any $\alpha >0$, otherwise ${\bf W}_1$ and ${\bf W}_2$ cannot be
%both optimal.
%
As a basic result in convex optimization,
any ${\bf W}_3=\beta {\bf W}_{1} + (1-\beta) {\bf
W}_{2}$, for $\beta \in (0,1)$, is also an optimal solution~\cite{Boyd2004}.
Since
${\bf W}_{1}$ and ${\bf W}_{2}$ are distinct, it can be easily shown
that ${\bf W}_3$ is of rank two, which violates the necessity that
any optimal ${\bf W}$ must be of rank one. In other words, we must
have one optimal ${\bf W}$ only.
%------------------------proof of Proposition 2----------------------

\subsection{Proof of Theorem~\ref{Theorem:SRM_rank1}} \label{appendix:Theorem1}
For ease of exposition, we restate Problems~\eqref{Relx_SRM_MISOMEs}
and \eqref{Relx_PM_SRC} in the following equations, respectively
\begin{equation}\label{eq:Thm1_rlx_SRM}
\begin{array}{rl}
\gamma^\star_{relax} =  \displaystyle \min_{{\bf W} \succeq {\bf 0}}
 & \phi ({\bf
W})\\
{\rm s.t.} &  {\rm Tr}({\bf W}) \le P
\end{array}
\end{equation}
and
\begin{equation}\label{eq:Thm1_rlx_SRC}
\begin{array}{rl}
\displaystyle \min_{{\bf W} \succeq {\bf 0} } &  {\rm Tr}({\bf W}) \\
{\rm s.t.} & \gamma^\star_{relax} \geq \phi ({\bf W})
\end{array}
\end{equation}
where, with a slight abuse of notations but for notational
simplicity, $\gamma_{relax}^{\star} (P)$ is replaced by
$\gamma_{relax}^{\star}$, and
\[ \phi ({\bf W}) = \displaystyle
\max_{k=1,\ldots,K} \frac{ 1 + {\rm Tr}({\bf G}_k^H {\bf W} {\bf
G}_k )}{ 1 + {\bf h}^H {\bf W} {\bf h}}\] is used to denote the
objective function of \eqref{eq:Thm1_rlx_SRM}.

Our proof is divided into three steps: First, we prove that an
optimal solution to Problem~\eqref{eq:Thm1_rlx_SRC} is also an
optimal solution to Problem~\eqref{eq:Thm1_rlx_SRM}; second, we
prove the converse; finally, we utilize Proposition~1 to establish
our claim in Theorem~1.

 {\bf Step~1}: Let $\bar{\bf W}$ be an optimal
solution of~\eqref{eq:Thm1_rlx_SRM}, and ${\hat {\bf W}}$ be an
optimal solution of~\eqref{eq:Thm1_rlx_SRC}. By noting that
$\bar{\bf W}$ is also feasible to~\eqref{eq:Thm1_rlx_SRC}, we have
that
\[ P \geq {\rm Tr}( \bar{\bf W} ) \geq {\rm Tr}( \hat{\bf W} ). \]
Hence, $\hat{\bf W}$ is also feasible to~\eqref{eq:Thm1_rlx_SRM}.
This further implies that $\phi(\hat{\bf W}) \ge
\gamma_{relax}^{\star}$. Moreover, as an optimal solution
of~\eqref{eq:Thm1_rlx_SRC}, $\hat{\bf W}$ must satisfy the
constraint in~\eqref{eq:Thm1_rlx_SRC}; i.e., $\gamma_{relax}^{\star}
\ge \phi(\hat{\bf W})$. We therefore have $\phi(\hat{\bf W}) =
\gamma_{relax}^{\star}$; in other words, $\hat{\bf W}$ is optimal to
\eqref{eq:Thm1_rlx_SRM}.

{\bf Step~2}: Suppose that $\bar{\bf W}$ is optimal to
Problem~\eqref{eq:Thm1_rlx_SRM}, but not optimal
to~\eqref{eq:Thm1_rlx_SRC}. Since $\bar{\bf W}$ is feasible
to~\eqref{eq:Thm1_rlx_SRC}, the following relation holds
\[{\rm Tr}(\hat{\bf W})< {\rm Tr}(\bar{\bf W}) \le P.\]
Since ${\rm Tr}(\hat{\bf W})<P$, we can construct another point
$\breve{\bf W}=\alpha_0 {\hat {\bf W}}$ with $\alpha_0>1$, such that
${\rm Tr}(\breve{\bf W})=P$; i.e., $\breve{\bf W}$ is feasible to
Problem~\eqref{eq:Thm1_rlx_SRM}. Now, consider the following
function
\begin{equation*}
\begin{aligned}
f(\alpha):= \phi ( \alpha \hat{\bf W}) = \frac{1+ \alpha
\displaystyle\max_{k=1,\ldots,K}{\rm Tr}({\bf G}_k^H {\hat {\bf W}}
{\bf G}_k)}{1+ \alpha {\bf h}^H {\hat {\bf W}} {\bf h}}.
\end{aligned}
\end{equation*}
The function $f(\alpha)$ is strictly decreasing with respect to
$\alpha$: It can be verified that
\begin{equation*}
\begin{aligned}
f'(\alpha)=\frac{\displaystyle\max_{k=1,\ldots,K} {\rm Tr}({\bf
G}_k^H {\hat {\bf W}} {\bf G}_k) - {\bf h}^H {\hat {\bf W}} {\bf h}}
{(1 + \alpha {\bf h}^H {\hat {\bf W}} {\bf h} )^2} < 0,
\end{aligned}
\end{equation*}
where the inequality above holds for ${\bf h}^H \hat{\bf W} {\bf
h} > \displaystyle\max_{k=1,\ldots,K} {\rm Tr}({\bf G}_k^H \hat{\bf
W} {\bf G}_k)$, which must be true for ${\gamma }_{relax}^{\star}
\le {\gamma}^{\star} < 1$. With the strictly decreasing property of
$f(\alpha)$ and $\alpha_0
>1$, we have that
\[\phi (\breve{\bf W}) = \phi (\alpha_0 \hat{\bf W}) = f(\alpha_0) < f (1) = \phi (\hat{\bf W}) = \gamma_{relax}^{\star}\]
which means that ${\breve {\bf W}}$ can achieve a lower objective
value than ${\hat {\bf W}}$ in Problem~\eqref{eq:Thm1_rlx_SRM}. This
contradicts the optimality of ${\hat {\bf W}}$ for
Problem~\eqref{eq:Thm1_rlx_SRM}.

{\bf Step~3}: So far, we have proven that
Problems~\eqref{eq:Thm1_rlx_SRM} and \eqref{eq:Thm1_rlx_SRC} have
the same optimal solution set. Since ${\gamma}_{relax}^{\star}<1$,
by Proposition~1, we know that there is a unique rank-one optimal
solution to Problem~\eqref{eq:Thm1_rlx_SRC}. Hence
Problem~\eqref{eq:Thm1_rlx_SRM} also admits a unique rank-one
optimal solution. Moreover, this rank-one optimal solution fulfills
the equality $\gamma^\star= {\gamma}_{relax}^{\star}$ (recall
Lemma~1), thereby serving as an optimal solution to the original
problem~\eqref{SRM_MISOMEs_reform} as well. On the other hand, let
${\bf W}^{\star}$ be an optimal solution of
\eqref{SRM_MISOMEs_reform}. Since ${\bf W}^{\star}$ is feasible to
Problem~\eqref{eq:Thm1_rlx_SRM} and the relaxation is tight, i.e.,
$\gamma^\star= {\gamma}_{relax}^{\star}$, ${\bf W}^{\star}$ is
optimal to Problem~\eqref{eq:Thm1_rlx_SRM}, too. This further
implies that ${\bf W}^{\star}$ has to be unique and of rank one.

%---------------------------------------------------------------------
\subsection{Proof of Proposition~\ref{prop:rob_PM_rank1}}\label{appendix:prop3}

The key to the proof lies in the KKT conditions. The Lagrangian
function of Problem~\eqref{Relx_rob_SRC_SDP} is given by
\begin{eqnarray}
\mathcal{L} (\bm{\mathcal{X}}) & = & {\rm
Tr}(\mathbf{W})-\sum_{k=1}^K{\rm Tr} \big( \mathbf{T}_{e,k}({\bf
W},\lambda_{e,k},
\theta)\mathbf{A}_{e,k} \big) \nonumber \\
& & -{\rm Tr} \big(\mathbf{T}_b({\bf W}, \lambda_b, \theta
)\mathbf{A}_b \big)-{\rm
Tr}(\mathbf{WS})-\lambda_b\mu_b-\sum_{k=1}^K\lambda_{e,k}\mu_{e,k}\label{robust_proof_Lagrangian}
\end{eqnarray}
where $\bm{\mathcal{X}} = \big\{ {\bf W}, {\bf S}, \lambda_b,
\bm{\lambda}_{e}, \theta, \mu_b, \bm{\mu}_{e}, \bm{ {\bf A}}_{e},
{\bf A}_b \big\}$
collects all the primal and dual variables, with
%{\color{blue} $\bm{\lambda}_e = \{ \lambda_{e,k} \}_{k=1}^{K}$,
% $\bm{\mu}_e = \{ \mu_{e,k} \}_{k=1}^{K}$,
$\bm{\lambda}_e= [~ \lambda_{e,1},\ldots,\lambda_{e,K} ~]^T$,
$\bm{\mu}_e = [~ \mu_{e,1},\ldots,\mu_{e,K} ~]^T$,
and $\bm{ {\bf A}}_{e} = \{ {\bf A}_{e,k}\}_{k=1}^{K}$;
%{\color{red} (it was
%$\bm{\mathcal{X}} = \{{\bf W}, {\bf T}_b, {\bf T}_e,
%\lambda_b, \bm{\lambda}_e, {\bf S}, {\bf A}_b, {\bf A}_e, \mu_b,
%\bm{\mu}_e \}$ )
%}
${\bf A}_{e,k}\in \mathbb{H}_{+}^{N_{e,k}N_t+1}, {\bf
A}_b\in\mathbb{H}_{+}^{N_t+1}, {\bf S}\in \mathbb{H}_{+}^{N_t},
\mu_b \in \mathbb{R}_+$ and $\mu_{e,k} \in \mathbb{R}_+$ are dual
variables associated with ${\bf T}_{e,k}$, ${\bf T}_b$, ${\bf W}$,
$\lambda_b$ and $\lambda_{e,k}$, respectively.

For ease of expression, we rewrite ${\bf T}_{b}({\bf W}, \lambda_b,
\theta)$ and ${\bf T}_{e,k} ({\bf W}, \lambda_{e,k}, \theta)$ as
\begin{eqnarray}
{\bf T}_{b}({\bf W}, \lambda_b, \theta) & = &
\bm{\Gamma}_{b}(\lambda_b, \theta) + {\bf V}_b^H {\bf W} {\bf V}_b, \label{proof_robust_LMI1}\\
{\bf T}_{e,k}({\bf W}, \lambda_{e,k}, \theta) & = &
\bm{\Gamma}_{e,k}(\lambda_{e,k}, \theta) - {\bf V}_{e,k}^H
\bm{\mathcal{W}}_k {\bf V}_{e,k},\label{proof_robust_LMI2}
\end{eqnarray}

where
\[ \bm{\Gamma}_b(\lambda_b, \theta) = \begin{bmatrix}
  \lambda_b {\bf I}_{N_t} & {\bf 0}\\
  {\bf 0} & -\lambda_b \varepsilon_b^2 - \theta +1
\end{bmatrix},\quad {\bf V}_b = [~ {\bf I}_{N_t}~ ~\bar{\bf h}~], \]
\[ \bm{\Gamma}_{e,k} (\lambda_{e,k}, \theta)= \begin{bmatrix}
  \lambda_{e,k} {\bf I}_{N_t N_{e,k}} & {\bf 0}\\
  {\bf 0} & -\lambda_{e,k} \varepsilon_{e,k}^2 + 2^{-R} \theta -1
\end{bmatrix}, \quad {\bf V}_{e,k} = [~ {\bf I}_{N_t N_{e,k}}~~ \bar{\bm
g}_k~]. \]
Substituting \eqref{proof_robust_LMI1} and \eqref{proof_robust_LMI2}
into \eqref{robust_proof_Lagrangian}, we obtain an alternate
expression of the Lagrangian function
  \begin{eqnarray}
\mathcal{L} (\bm{\mathcal{X}}) & = & {\rm Tr}(\mathbf{W})+
\sum_{k=1}^{K} {\rm Tr}\big(\bm{\mathcal{W}}_k {\bf V}_{e,k} {\bf
A}_{e,k} {\bf V}_{e,k}^H \big) - {\rm Tr} \big( {\bf W} {\bf V}_b
{\bf A}_{b} {\bf V}_b^H
\big) \nonumber\\
 & & - {\rm
Tr}(\mathbf{WS}) + \varphi \big(\lambda_b, \bm{\lambda}_{e}, \theta,
\mu_b, \bm{\mu}_{e}, \bm{ {\bf A}}_{e},
{\bf A}_b \big),\nonumber\\
&  = & {\rm Tr}(\mathbf{W})+ \sum_{k=1}^{K} \sum_{l=1}^{N_{e,k}}
{\rm Tr}\big( {\bf W} {\bf B}_{e,k}^{(l,l)} \big) - {\rm Tr} \big(
{\bf W} {\bf V}_b {\bf A}_{b} {\bf V}_b^H
\big) \nonumber\\
& & - {\rm Tr}(\mathbf{WS}) + \varphi \big(\lambda_b,
\bm{\lambda}_{e}, \theta, \mu_b, \bm{\mu}_{e}, \bm{ {\bf A}}_{e},
{\bf A}_b \big),\label{robust_proof_lagrangian_2}
\end{eqnarray}
where ${\bf B}_{e,k}^{(l,l)} \in \mathbb{H}_{+}^{N_t}$ is a block
submatrix of ${\bf V}_{e,k} {\bf A}_{e,k} {\bf V}_{e,k}^H$; specifically,
\[{\bf V}_{e,k} {\bf A}_{e,k} {\bf V}_{e,k}^H = \left[
                                                  \begin{array}{lcl}
                                                    {\bf B}_{e,k}^{(1,1)} & \ldots & {\bf B}_{e,k}^{(1,N_{e,k})} \\
                                                    \vdots & \ddots & \vdots \\
                                                    {\bf B}_{e,k}^{(N_{e,k},1)} & \ldots & {\bf B}_{e,k}^{(N_{e,k},N_{e,k})} \\
                                                  \end{array}
                                                \right] \in \mathbb{H}_{+}^{N_tN_{e,k}};
\]
and $ \varphi \big(\lambda_b, \bm{\lambda}_{e}, \theta, \mu_b,
\bm{\mu}_{e}, \bm{ {\bf A}}_{e}, {\bf A}_b \big) $ collects the
terms not related to ${\bf W}$ and ${\bf S}$, which are not
important to the proof.

We consider only the KKT conditions relevant to the proof here:
\begin{subequations}\label{eq:rob_pow_min_KKT}
\begin{align}
\nabla_{\bf W} \mathcal{L}(\bm{\mathcal{X}}) & = {\bf 0},\label{eq:rob_pow_min_KKT_a0}\\
\mathbf{T}_b({\bf W}, \lambda_b, \theta) \mathbf{A}_b&=\mathbf{0},\label{eq:rob_pow_min_KKT_b}\\
\mathbf{WS}&={\bf 0},\label{eq:rob_pow_min_KKT_c}\\
\lambda_b \ge 0, ~ {\bf W} \succeq {\bf 0}, ~ {\bf A}_b \succeq {\bf
0}, ~ {\bf A}_{e,k} & \succeq {\bf
0},~~k=1,\ldots,K.\label{eq:rob_pow_min_KKT_d}
\end{align}
\end{subequations}
Using the expression \eqref{robust_proof_lagrangian_2}, the KKT
condition \eqref{eq:rob_pow_min_KKT_a0} is obtained as follows:
\begin{equation}\label{eq:rob_pow_min_KKT_a}
\mathbf{I}_{N_t}+\sum_{k=1}^{K}\sum_{l=1}^{N_{e,k}}{\bf
B}_{e,k}^{(l,l)} - {\bf V}_b {\bf A}_b {\bf V}_b^H - {\bf S} = {\bf
0}.
\end{equation}
Premultiplying the two sides of \eqref{eq:rob_pow_min_KKT_a} by
${\bf W}$, and making use of \eqref{eq:rob_pow_min_KKT_c}, we get
\begin{align}
\mathbf{W}\Bigl({\bf
I}_{N_{t}}+\sum_{k=1}^{K}\sum_{l=1}^{N_{e,k}}{\bf
B}_{e,k}^{(l,l)}\Bigr)=\mathbf{W}{\bf V}_b {\bf A}_b {\bf
V}_b^H.\label{eq:rob_pow_min_KKT_f}
\end{align}
Now the following relation holds:
\begin{subequations}\label{proof_robust_rank_1}
\begin{align}
 {\rm rank}({\bf W}) & =  {\rm rank} \left(\mathbf{W}\Bigl({\bf
I}_{N_{t}}+\sum_{k=1}^{K}\sum_{l=1}^{N_{e,k}}{\bf B}_{e,k}^{(l,l)}\Bigr) \right) \label{proof_robust_rank_1_a}\\
& =  {\rm rank}(\mathbf{W}{\bf V}_b {\bf A}_b {\bf V}_b^H ) \label{proof_robust_rank_1_b} \\
& \le  \min \{ {\rm rank}({\bf V}_b {\bf A}_b {\bf V}_b^H),~{\rm
rank}({\bf W}) \}\label{proof_robust_rank_1_c}
\end{align}
\end{subequations}
where \eqref{proof_robust_rank_1_a} is due to ${\bf
I}_{N_{t}}+\sum_{k=1}^{K}\sum_{l=1}^{N_{e,k}}{\bf B}_{e,k}^{(l,l)}
\succ {\bf 0}$, \eqref{proof_robust_rank_1_b} and
\eqref{proof_robust_rank_1_c} follow from
\eqref{eq:rob_pow_min_KKT_f} and a basic rank inequality property~\cite{Horn85}. If we can prove that ${\rm rank}({\bf
V}_b {\bf A}_b {\bf V}_b^H) = 1$, then, from
\eqref{proof_robust_rank_1}, we will obtain ${\rm rank}({\bf W}) \le
1$. Therefore, in the remaining part of the proof, we will focus on
the rank of ${\bf V}_b {\bf A}_b {\bf V}_b^H $.

Substituting \eqref{proof_robust_LMI1} into the KKT condition
\eqref{eq:rob_pow_min_KKT_b}, we obtain
\begin{equation}\label{robust_proof_complement_a}
\bm{\Gamma}_{b}(\lambda_b, \theta) {\bf A}_b + {\bf V}_b^H {\bf W}
{\bf V}_b {\bf A}_b =  {\bf 0}.
\end{equation}
And it follows by postmultiplying \eqref{robust_proof_complement_a}
by ${\bf V}_b^H$ that
\begin{equation}\label{robust_proof_complement_b}
\bm{\Gamma}_{b}(\lambda_b, \theta) {\bf A}_b {\bf V}_b^H + {\bf
V}_b^H {\bf W} {\bf V}_b {\bf A}_b {\bf V}_b^H   =  {\bf 0}.
\end{equation}
By noting the following facts
\begin{eqnarray}\label{robust_proof_facts}
~[~ {\bf I}_{N_t}~ ~ {\bf 0} ~] \bm{\Gamma}_{b}(\lambda_b, \theta) &
= & \lambda_b [~{\bf I}_{N_t} ~ ~ {\bf 0} ~ ] = \lambda_b({\bf V}_b
- [~ {\bf 0}_{N_t} ~~
\bar{\bf h}~]), \nonumber\\
~[~ {\bf I}_{N_t}~ ~ {\bf 0} ~] {\bf V}_b^H & = & {\bf I}_{N_t},
\end{eqnarray}
we premultiply the both sides of \eqref{robust_proof_complement_b}
by $[~ {\bf I}_{N_t}~~ {\bf 0}~]$ to get
\begin{subequations} \label{robust_proof_complement2}
\begin{align}
& \lambda_b({\bf V}_b - [~ {\bf 0}_{N_t} ~~ \bar{\bf h}~]) {\bf A}_b
{\bf V}_b^H + {\bf W} {\bf V}_b {\bf A}_b {\bf
V}_b^H  = {\bf 0}, \label{robust_proof_complement2a}\\
\Leftrightarrow ~ &  (\lambda_b {\bf I}_{N_t} + {\bf W}) {\bf V}_b
{\bf A}_b {\bf V}_b^H = \lambda_b[~ {\bf 0}_{N_t} ~~ \bar{\bf h}~]
{\bf A}_b {\bf V}_b^H. \label{robust_proof_complement2c}
\end{align}
\end{subequations}

We claim that $\lambda_b$ must be positive. Suppose that
$\lambda_b=0$. Then, according to \eqref{robust_proof_complement2a},
we have ${\bf W} {\bf V}_b {\bf A}_b {\bf V}_b^H  = {\bf 0}$. By
\eqref{eq:rob_pow_min_KKT_f} and ${\bf
I}_{N_{t}}+\sum_{k=1}^{K}\sum_{l=1}^{N_{e,k}}{\bf B}_{e,k}^{(l,l)}
\succ {\bf 0}$, we have $\mathbf{W}=\mathbf{0}$. However, ${\bf W}=
{\bf 0}$ is infeasible to the relaxed R-SRC problem
\eqref{Relx_rob_SRC}, whenever $R>0$. Therefore, $\lambda_b>0$ must
hold. With $\lambda_b >0$, we have that
\begin{subequations}\label{eq:rank_relation}
\begin{align}
{\rm rank}({\bf V}_b {\bf A}_b {\bf V}_b^H) & = {\rm rank} \big(
(\lambda_b {\bf I}_{N_t} + {\bf W}) {\bf V}_b {\bf A}_b {\bf
V}_b^H \big)\label{eq:rank_relation_a} \\
& = {\rm rank} \big(\lambda_b[~ {\bf 0}_{N_t} ~~ \bar{\bf h}~] {\bf
A}_b
{\bf V}_b^H \big) \label{eq:rank_relation_b}\\
& \le {\rm rank} \big([~ {\bf 0}_{N_t} ~~ \bar{\bf h}~]\big) \le 1,
\label{eq:rank_relation_c}
\end{align}
\end{subequations}
where \eqref{eq:rank_relation_a} is due to $\lambda_b {\bf I}_{N_t}
+ {\bf W} \succ {\bf 0}$, \eqref{eq:rank_relation_b} and
\eqref{eq:rank_relation_c} follow from
\eqref{robust_proof_complement2c} and a basic rank inequality
property~\cite{Horn85}.

Combining \eqref{proof_robust_rank_1} and \eqref{eq:rank_relation},
we have
\[ {\rm rank}({\bf W}) \le {\rm rank}({\bf V}_b {\bf A}_b {\bf V}_b^H) \le
1.\] Since $\mathbf{W}\ne \mathbf{0}$ for $R>0$, the rank of
$\mathbf{W}$ must be one.

Regarding the uniqueness of the optimal solution, the proof is
exactly the same as that in Proposition~1. We therefore omit it for
brevity.

%-----------------proof of proposition 4---SDP reformulation of R-SRM------------
\subsection{Proof of Proposition~\ref{prop:R-SRM_reformulation}}\label{appendix:prop4}
By the change of variable ${\bf W} = {\bf Z}/ \xi$, $\xi > 0$,
Problem~\eqref{Relx_rob_SRM} can be transformed to
\begin{subequations}\label{Relx_rob_SRM_reform}
\begin{align}
\min_{ {\bf Z}, \xi } & ~ \frac{ \xi + \displaystyle \max_{k=1,\ldots,K} \max_{ {\bf G}_k \in \mathcal{B}_{e,k} } {\rm Tr}( {\bf G}_k^H {\bf Z} {\bf G}_k ) }{ \xi +  \displaystyle \min_{ {\bf h} \in \mathcal{B}_b } {\bf h}^H {\bf Z} {\bf h}  } \\
{\rm s.t.} & ~ {\rm Tr}({\bf Z}) \leq \xi P, \\
& ~ {\bf Z} \succeq {\bf 0}, ~ \xi > 0.
\end{align}
\end{subequations}
We first show that~\eqref{Relx_rob_SRM_reform} is equivalent to the
following problem
\begin{subequations}\label{Relx_rob_SRM_reform_2}
\begin{align}
\min_{ {\bf Z}, \xi } & ~  \xi + \displaystyle \max_{k=1,\ldots,K} \max_{ {\bf G}_k \in \mathcal{B}_{e,k}} {\rm Tr}( {\bf G}_k^H {\bf Z} {\bf G}_k )  \label{Relx_rob_SRM_reform_2a}\\
{\rm s.t.} & ~ \xi +  \displaystyle \min_{ {\bf h} \in \mathcal{B}_b } {\bf h}^H {\bf Z} {\bf h} \geq 1, \label{Relx_rob_SRM_reform_2b}\\
& ~ {\rm Tr}({\bf Z}) \leq \xi P, \label{Relx_rob_SRM_reform_2c}\\
& ~ {\bf Z} \succeq {\bf 0}, ~ \xi \geq
0.\label{Relx_rob_SRM_reform_2d}
\end{align}
\end{subequations}
Consider the optimal solution of~\eqref{Relx_rob_SRM_reform_2}, say,
denoted by $({\bf Z}^\star, \xi^\star)$.
From~\eqref{Relx_rob_SRM_reform_2b}
and~\eqref{Relx_rob_SRM_reform_2c}, it can be verified that
$\xi^\star > 0$ must hold. Hence, $({\bf Z}^\star, \xi^\star)$ is
feasible to~\eqref{Relx_rob_SRM_reform}. One can deduce that $({\bf
Z}^\star, \xi^\star)$ is optimal to~\eqref{Relx_rob_SRM_reform} if
it holds true that
\[ \xi^\star +   \min_{ {\bf h} \in \mathcal{B}_b } {\bf h}^H {\bf Z}^\star {\bf h} =  1. \]
We use contradiction to verify the latter. Suppose that $\xi^\star +
\min_{ {\bf h} \in \mathcal{B}_b } {\bf h}^H {\bf Z}^\star {\bf h} >
1.$ Then we can choose a feasible point $( \tilde{\bf Z},
\tilde{\xi} ) = (\alpha{\bf Z}^\star, \alpha\xi^\star)$ for some $0<
\alpha < 1$ such that $\tilde{\xi} +   \min_{ {\bf h} \in
\mathcal{B}_b } {\bf h}^H \tilde{\bf Z} {\bf h} = 1.$ The point $(
\tilde{\bf Z}, \tilde{\xi} )$ can be verified to achieve an
objective value lower than that offered by the optimal point $({\bf
Z}^\star, \xi^\star)$, which is a contradiction.

Our next step is to turn \eqref{Relx_rob_SRM_reform_2} to an SDP.
Using the epigraph reformulation, \eqref{Relx_rob_SRM_reform_2} can
be rewritten as
\begin{subequations}\label{Relx_rob_SRM_reform_3}
\begin{align}
\min_{ {\bf Z}, \xi, \tau } & ~  \tau  \label{Relx_rob_SRM_reform_3a}\\
{\rm s.t.} & ~ \tau \geq \xi + \displaystyle \max_{ {\bf G}_k \in \mathcal{B}_{e,k}} {\rm Tr}( {\bf G}_k^H {\bf Z} {\bf G}_k ), ~ k=1,\ldots,K, \label{Relx_rob_SRM_reform_3b}\\
& ~ \xi +  \displaystyle \min_{ {\bf h} \in \mathcal{B}_b } {\bf h}^H {\bf Z} {\bf h} \geq 1, \label{Relx_rob_SRM_reform_3c}\\
& ~ {\rm Tr}({\bf Z}) \leq \xi P, \label{Relx_rob_SRM_reform_3d}\\
& ~ {\bf Z} \succeq {\bf 0}, ~ \xi \geq
0.\label{Relx_rob_SRM_reform_3e}
\end{align}
\end{subequations}
By applying the $\mathcal{S}$-procedure to convert the
constraints~\eqref{Relx_rob_SRM_reform_3b}
and~\eqref{Relx_rob_SRM_reform_3c} into LMIs, we obtain the
SDP~\eqref{Relx_rob_SRM_SDP}.

%---------------------------------------------------------------------
\subsection{Worst-case Secrecy Rate Calculation}\label{appendix:cal_worst_sec}

The worst-case secrecy rate function $\psi({\bf W})$ in
\eqref{rob_SRM_part}-\eqref{rob_SRM_part_end} can be computed when
${\bf W}$ is of rank one (which is the case of all the considered
methods in this paper). With rank-one ${\bf W}$, $\psi ({\bf W})$
can be reduced to
\begin{equation}\label{eq:worst_sec_nonrobust_main}
\psi({\bf W})=\min_{{\bf h} \in \mathcal{B}_b}\log(1+{\bf h}^H {\bf
W } {\bf h})-\max_{k=1,\ldots,K}\max_{{\bf G}_k \in
\mathcal{B}_{e,k}}\log(1+{\rm Tr}({\bf G}_k^H {\bf W} {\bf G}_k)).
\end{equation}
The first and second terms of \eqref{eq:worst_sec_nonrobust_main}
are separate optimization problems. They can be recast as the
following two SDPs by using the ${\cal S}$-procedure:
\begin{equation}\label{eq:worst_sec_nonrobust_subproblems_1}
\begin{aligned}
\tau_1^{\star} = \max_{\tau_1,\lambda_b} & ~ \tau_1 \notag\\
{\rm s.t.} & ~ \begin{bmatrix} \lambda_b {\bf I}_{N_t}+{\bf W} &
{\bf W} {\bf \bar{h}}
\\ {\bf \bar{h}}^H {\bf W} & {\bf \bar{h}}^H {\bf W} {\bf
\bar{h}}+1-\tau_1-\lambda_b\varepsilon_b^2\end{bmatrix} \succeq {\bf 0},\notag\\
& ~ \lambda_b\ge 0,
\end{aligned}
\end{equation}
and
\begin{equation}\label{eq:worst_sec_nonrobust_subproblems_2}
\begin{aligned}
\tau_2^{\star} = \min_{\tau_2,
%{\color{blue} \{ \lambda_{e,k}\}_{k=1}^{K}}
\lambda_{e,1},\ldots,\lambda_{e,K}
} & ~ \tau_2 \notag \\
{\rm s.t.} & ~
\begin{bmatrix}\lambda_{e,k}\mathbf{I}_{N_{e,k} N_t}-\bm{\mathcal{W}}_{k}
& -\bm{ \mathcal{W} }_{k} \bm{\bar{g}}_k\\
-\bm{\bar{g}}_k^H \bm{ \mathcal{W}}_{k} &
-\lambda_{e,k}\varepsilon_{e,k}^2 - \bm{\bar{g}}_k^H \bm{ \mathcal{W} }_{k} \bm{\bar{g}}_k + \tau_2-1\end{bmatrix} \succeq {\bf 0}, \notag \\
& ~ \lambda_{e,k}\ge 0,  ~ k=1,\ldots,K,
\end{aligned}
\end{equation}
where $\bm{\mathcal{W}}_{k} = {\bf I}_{N_{e,k}} \otimes {\bf W} $.
Once the optimal values $\tau_1^{\star}$ and $\tau_2^{\star}$ are computed (by an available SDP solver), the
 worst-case secrecy rate is obtained as
\[\psi ({\bf W})=\log(\tau_1^{\star}) - \log(\tau_2^{\star}).\]

\end{document}